\documentclass[twocolumn]{aastex63}

\shorttitle{The MZR at $z\sim2.2$ in overdense environments.}
\shortauthors{Wang et al. (2021)}

\hypersetup{linkcolor=dgreen,citecolor=lblue,filecolor=cyan,urlcolor=magenta}

\usepackage[english]{babel} 
\usepackage[utf8]{inputenc} 
\usepackage[T1]{fontenc}    
\DeclareUnicodeCharacter{0301}{/}  
\DeclareUnicodeCharacter{2212}{-}  
\usepackage{ae,aecompl}
\usepackage{pgf,pgfarrows,pgfnodes,pgfautomata,pgfheaps}
\usepackage{graphicx}       
\usepackage{natbib}         
\usepackage{url}            
\usepackage{grffile}        
\usepackage{mathtools}      
\usepackage{multirow}       
\usepackage{xspace}         

\usepackage{txfonts}
\usepackage{amsmath,amssymb,amsxtra,amsfonts}   

\newcommand{\lya}{\textrm{Ly}\ensuremath{\alpha}\xspace}

\usepackage{color}         
\definecolor{gold}{rgb}{1,0.80,0}
\definecolor{orange}{rgb}{1,0.5,0}
\definecolor{midgray}{gray}{0.3}
\definecolor{lblue}{rgb}{0,0.2,0.6}
\definecolor{dgreen}{rgb}{0.1,0.6,0.3}
\definecolor{purple}{rgb}{0.5019607843137255,0.0,0.5019607843137255}

\renewcommand\farcs{\mbox{$.\!^{\prime\prime}$}}    
\renewcommand{\arcsec}{\mbox{$^{\prime\prime}$}\xspace}  
\renewcommand{\arcdeg}{\mbox{$^{\circ}$}\xspace}             

\newcommand{\be}{\begin{equation}}
\newcommand{\ee}{\end{equation}}

\newcommand{\ba}{\begin{align}}
\newcommand{\ea}{\end{align}}

\newcommand{\defeq}{\vcentcolon=}

\newcommand{\Msun}{\ensuremath{M_\odot}\xspace}

\newcommand{\Mstar}{\ensuremath{M_\ast}\xspace}

\newcommand{\oh}{\ensuremath{12+\log({\rm O/H})}\xspace}

\newcommand{\K}{\ensuremath{\rm K}\xspace}

\newcommand{\Hunit}{\ensuremath{\rm km~s^{-1}~Mpc^{-1}}\xspace}
\newcommand{\Funit}{\ensuremath{\rm erg~s^{-1}~cm^{-2}}\xspace}

\def\micron{\ensuremath{\mu\textrm{m}}\xspace}  

\newcommand\ionp[2]{#1$\;${\scshape{#2}}}      
\newcommand{\Ha}{\textrm{H}\ensuremath{\alpha}\xspace}
\newcommand{\Hb}{\textrm{H}\ensuremath{\beta}\xspace}
\newcommand{\Hg}{\textrm{H}\ensuremath{\gamma}\xspace}
\newcommand{\Hd}{\textrm{H}\ensuremath{\delta}\xspace}
\newcommand{\HII}{\textrm{H}\textsc{ii}\xspace}

\newcommand{\OI}{[\textrm{O}~\textsc{i}]\xspace}
\newcommand{\OII}{[\textrm{O}~\textsc{ii}]\xspace}
\newcommand{\OIII}{[\textrm{O}~\textsc{iii}]\xspace}

\newcommand{\NII}{[\textrm{N}~\textsc{ii}]\xspace}
\newcommand{\SII}{[\textrm{S}~\textsc{ii}]\xspace}
\newcommand{\NeIII}{[\textrm{Ne}~\textsc{iii}]\xspace}

\def\U{\ensuremath{U_{360}}\xspace}     
\def\B{\ensuremath{B_{435}}\xspace}
\def\V{\ensuremath{V_{606}}\xspace}
\def\I{\ensuremath{I_{814}}\xspace}
\def\Y{\ensuremath{Y_{105}}\xspace}
\def\J{\ensuremath{J_{125}}\xspace}

\def\H{\ensuremath{H_{160}}\xspace}

\newcommand{\sex}{\textsc{SExtractor}\xspace}
\newcommand{\emc}{\textsc{Emcee}\xspace}

\newcommand{\adriz}{\textsc{AstroDrizzle}\xspace}

\newcommand{\grzl}{\textsc{Grizli}\xspace}

\newcommand{\tphot}{\textsc{T-PHOT}\xspace}
\newcommand{\bagp}{\textsc{BAGPIPES}\xspace}

\newcommand{\hst}{\textit{HST}\xspace}

\newcommand{\mg}{\textit{MAMMOTH-Grism}\xspace}

\def\ie{i.e.\xspace}
\def\eg{e.g.\xspace}

\renewcommand\({\left(}
\renewcommand\){\right)}

\newcommand{\el}[1]{\ensuremath{\textrm{EL}_{#1}}}

\def\p{{\rm prior}}

\newcommand{\Om} {\ensuremath{\Omega_{\rm{m}}}\xspace}

\newcommand{\Ol} {\ensuremath{\Omega_{\Lambda}}\xspace}

\newcommand{\n}{\ensuremath{{\nu}\rm}\xspace}

\usepackage{etoolbox}
\makeatletter
\patchcmd{\NAT@citex}
  {\@citea\NAT@hyper@{\NAT@nmfmt{\NAT@nm}\NAT@date}}
  {\@citea\NAT@nmfmt{\NAT@nm}\NAT@hyper@{\NAT@date}}
  {}
  {}
\patchcmd{\NAT@citex}
  {\@citea\NAT@hyper@{%
     \NAT@nmfmt{\NAT@nm}%
     \hyper@natlinkbreak{\NAT@aysep\NAT@spacechar}{\@citeb\@extra@b@citeb}%
     \NAT@date}}
  {\@citea\NAT@nmfmt{\NAT@nm}%
   \NAT@aysep\NAT@spacechar%
   \NAT@hyper@{\NAT@date}}
  {}
  {}
\patchcmd{\NAT@citex}
  {\@citea\NAT@hyper@{%
     \NAT@nmfmt{\NAT@nm}%
     \hyper@natlinkbreak{\NAT@spacechar\NAT@@open\if*#1*\else#1\NAT@spacechar\fi}%
       {\@citeb\@extra@b@citeb}%
     \NAT@date}}
  {\@citea\NAT@nmfmt{\NAT@nm}%
   \NAT@spacechar\NAT@@open\if*#1*\else#1\NAT@spacechar\fi%
   \NAT@hyper@{\NAT@date}}
  {}
  {}
\makeatother

\usepackage{enumitem}
\setenumerate{labelindent=0.3cm,labelsep=0.1cm,leftmargin=*,itemsep=0.1cm,topsep=0.15cm,parsep=1pt,partopsep=0pt}

\begin{document}


\title{The mass-metallicity relation at cosmic noon in overdense environments: first results from the \mg \hst slitless spectroscopic survey}


\author{Xin Wang}
\affil{Infrared Processing and Analysis Center, Caltech, 1200 E. California Blvd., Pasadena, CA 91125, USA; wangxin@ipac.caltech.edu}

\author{Zihao Li}
\affil{Department of Astronomy and Tsinghua Center for Astrophysics, Tsinghua University, Beijing 100084, China}

\author{Zheng Cai}
\affil{Department of Astronomy and Tsinghua Center for Astrophysics, Tsinghua University, Beijing 100084, China; zcai@mail.tsinghua.edu.cn }

\author{Dong Dong Shi}
\affil{Purple Mountain Observatory, Chinese Academy of Sciences, 10 Yuan Hua Road, Nanjing 210023, China; ddshi@pmo.ac.cn}

\author{Xiaohui Fan}
\affil{Steward Observatory, University of Arizona, 933 North Cherry Avenue, Tucson, AZ 85721, USA}

\author{Xian~Zhong Zheng}
\affil{Purple Mountain Observatory, Chinese Academy of Sciences, 10 Yuan Hua Road, Nanjing 210023, China}

\author{Fuyan Bian}
\affil{European Southern Observatory, Alonso de Córdova 3107, Casilla 19001, Vitacura, Santiago 19, Chile}

\author{Harry I. Teplitz}
\affil{Infrared Processing and Analysis Center, Caltech, 1200 E. California Blvd., Pasadena, CA 91125, USA}

\author{Anahita Alavi}
\affil{Infrared Processing and Analysis Center, Caltech, 1200 E. California Blvd., Pasadena, CA 91125, USA}

\author{James Colbert}
\affil{Infrared Processing and Analysis Center, Caltech, 1200 E. California Blvd., Pasadena, CA 91125, USA}

\author{Alaina L. Henry}
\affil{Space Telescope Science Institute, 3700 San Martin Drive, Baltimore, MD, 21218, USA}

\author{Matthew~A.~Malkan}
\affil{Department of Physics and Astronomy, University of California, Los Angeles, CA 90095-1547, USA}

\begin{abstract}
    The \mg slitless spectroscopic survey is a Hubble Space Telescope (\hst) cycle-28 medium program, which is obtaining 45 orbits of WFC3/IR grism spectroscopy in the density peak regions of three massive galaxy protoclusters at $z=2-3$ discovered using the MAMMOTH technique. We introduce this survey by presenting the first measurement of the mass-metallicity relation (MZR) at high redshift in overdense environments via grism spectroscopy.
    From the completed \mg observations in the field of the BOSS1244 protocluster at $z=2.24\pm0.02$,
    we secure a sample of 36 protocluster member galaxies at $z\approx2.24$, showing strong nebular emission lines (\OIII, \Hb and \OII) in their G141 spectra.
    Using the multi-wavelength broad-band deep imaging from \hst and ground-based telescopes,
    we measure their stellar masses in the range of $[10^{9}, 10^{10.4}]$\,\Msun, instantaneous star formation rates (SFR) from 10 to 240\,\Msun\,yr$^{-1}$, and global gas-phase metallicities $[\frac{1}{3}, 1]$ of solar.
    Compared with similarly selected field galaxy sample at the same redshift, our galaxies show on average increased SFRs by $\sim$0.06\,dex and $\sim$0.18\,dex at $\sim$10$^{10.1}$\Msun and $\sim$10$^{9.8}$\Msun, respectively.
    Using the stacked spectra of our sample galaxies, we derive the MZR in the BOSS1244 protocluster core as $\oh = \left(0.136\pm0.018\right) \times \log(\Mstar/\Msun) + \left(7.082 \pm 0.175 \right)$, showing significantly shallower slope than that in the field.
    This shallow MZR slope is likely caused by the combined effects of efficient recycling of feedback-driven winds and cold-mode gas accretion in protocluster environments.
    The former effect helps low-mass galaxies residing in overdensities retain their metal production, 
    whereas the latter effect dilutes the metal content of high-mass galaxies, making them more metal poor than their coeval field counterparts.
\end{abstract}


\section{Introduction}\label{sect:intro}

Mapping galaxy properties in different environments across cosmic time is essential to forming a complete picture of galaxy formation and evolution.
Galaxy protoclusters at $z=2-3$ 
provide a direct probe of the rapid assembly of large-scale structures 
at cosmic noon, making them ideal laboratories to test the environmental dependence of galaxy mass assembly and chemical enrichment at the peak of cosmic star formation \citep[e.g.,][]{Overzier:2016iz,Shimakawa:2018ft}.
The relative abundance of oxygen compared to hydrogen in the interstellar medium --- gas-phase metallicity (hereafter referred to as metallicity for simplicity) --- provides a crucial diagnostic of the past history of star formation and complex gas movements driven by galactic feedback and tidal interactions \citep{Lilly:2013ko,Maiolino:2019vq}.
Therefore, measuring the metallicity of galaxies undergoing active star formation in the cores of protoclusters at cosmic noon can probe the effects of galactic feedback versus environment in shaping galaxy chemical evolution \citep{Oppenheimer:2008bu,Kulas:2013ica,Ma:2016gw}.

It is well established that metallicity correlates strongly with galaxy stellar mass (\ie the mass-metallicity relation: MZR) in both the local and distant universe \citep{Tremonti:2004ed,Erb:2006kn,2008A&A...488..463M,Andrews:2013dn,Henry:2021ju,Sanders:2021ga}.
There is also growing consensus about the role that environment plays in galaxy metal enrichment at $z\sim0$: galaxies situated in overdensities are observed to be more metal enriched than coeval field galaxies of the same mass \citep{Cooper:2008ge,Ellison:2009ch}.
However, at cosmic noon, the situation is highly uncertain, with little or conflicting evidence of the existence of any environmental effects \citep{Kacprzak:2015cg,Valentino:2015gv,Shimakawa:2015iwa}, primarily due to small number statistics.


In this paper, we present the first attempt to measure the high-$z$ MZR in overdense environments using space-based slitless spectroscopy, from a sample of 36 protocluster member galaxies showing prominent nebular emission lines.
The high throughput of the Wide-Field Camera 3 (WFC3) grism channels onboard the Hubble Space Telescope (\hst) and the low infrared (IR) sky background in space enable us to probe the metal enrichment in protocluster member galaxies down to very low stellar mass of $\Mstar\sim10^9\,\Msun$ at $z=2.24$.
In addition, we measure a pure emission-line selected sample, unbiased by photometric preselection, due to the slitless nature of WFC3's grism spectroscopy.
Throughout this work, we adopt the AB magnitude system and the standard concordance cosmology ($\Om=0.3, \Ol=0.7$, $H_0=70\,\Hunit$).  The metallic lines are denoted in the following manner, if presented without wavelength: $\OIII\lambda5008\defeq\OIII$, $\OII\lambda\lambda3727,3730\defeq\OII$, $\NeIII~3869\defeq\NeIII$, $\NII~\lambda6585\defeq\NII$.

\section{Grism observation of the MAMMOTH protocluster BOSS1244}\label{sect:obs}

\begin{figure*}[!htb]
    \centering
    \includegraphics[width=\textwidth, trim = 0cm 0cm 0cm 0cm, clip]{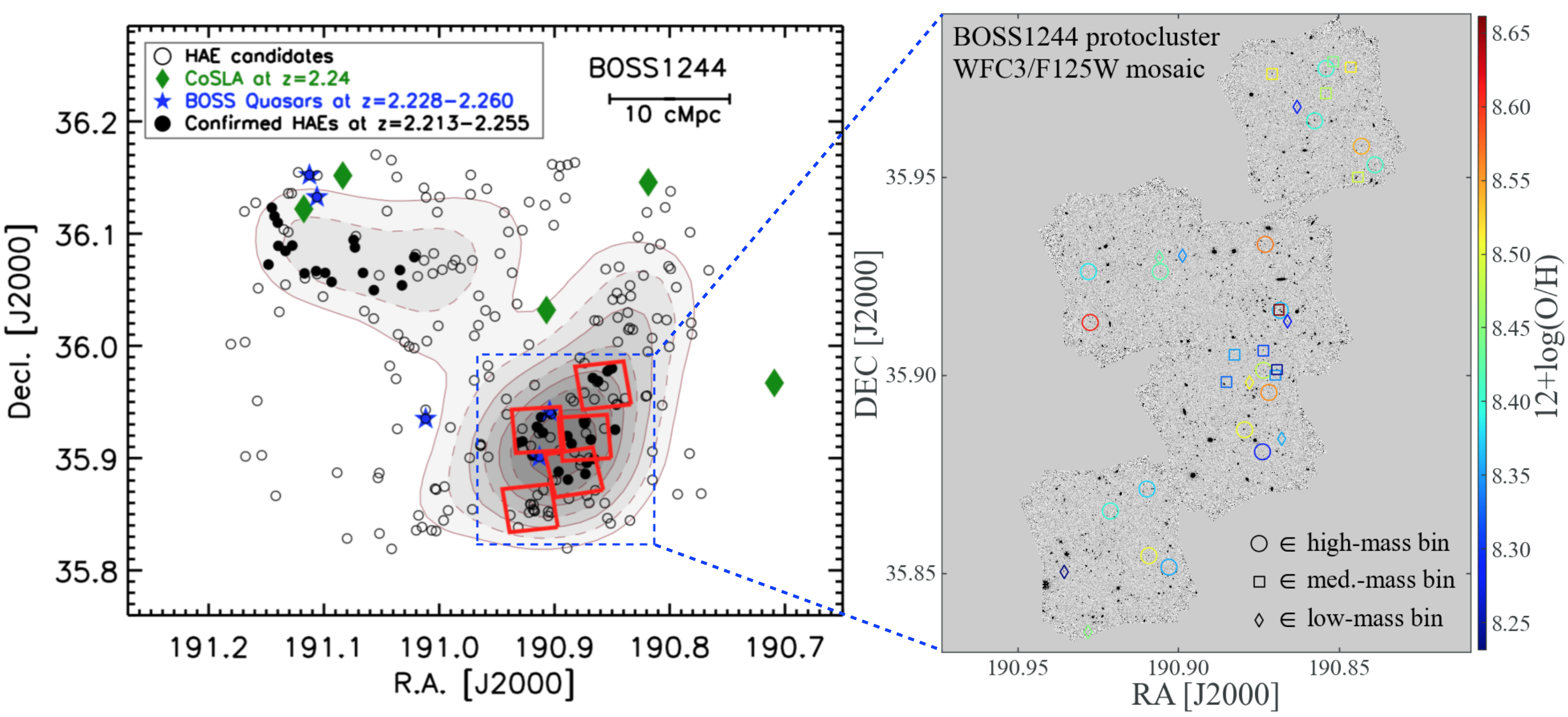}
    \caption{The BOSS1244 protocluster center field targeted by the \mg program and presented in this work.
    {\bf Left}: the density map for the massive protocluster BOSS1244 at $z=2.24\pm0.02$. The density contours (shown in brown and filled in grey) for \Ha emitters (HAEs) are shown in steps of 0.2\,arcmin$^{-2}$, with the inner density peak reaching $\sim$2\,arcmin$^{-2}$.
    The green diamonds are the groups of Ly$\alpha$ absorption systems from background high-$z$ BOSS quasars,
    and the blue stars are the quasars at the protocluster redshift.
    The black circles are the HAE candidates selected through CFHT broad+narrow band imaging and the black solid points are spectroscopically confirmed by ground-based IR spectroscopy \citep{Shi:2021jt}. It is expected to have a present-day total enclosed mass of $\gtrsim1.6\times10^{15}$\,\Msun, likely evolving into a local Coma-like supercluster.
    The density peak regions of the BOSS1244 protocluster in the southwest are targeted by \mg, with 5 pointings of WFC3/G141+F125W, each pointing at 3-orbit depth, highlighted by the red squares.
    {\bf Right}:
    A zoom-in view of the BOSS1244 density peak regions (marked by blue dashed box in the left panel) on the F125W image mosaics acquired by \mg. This F125W imaging is a coaddition of the pre-imaging component (with an average depth of 1800 sec) to the G141 grism exposures. Using the deep G141 grism spectroscopy, we secure a total of 36 star-forming protocluster member galaxies at $z\sim2.24$, showing extreme emission lines of \OIII, \Hb, and \OII. They are color-coded by their inferred metallicity assuming the \citet[][B18]{Bian:2018km} strong line calibrations as described in Sect.~\ref{subsect:oh12}, and represented by three sets of symbols corresponding to the three mass bins, chosen for the stacking analysis (see Sect.~\ref{subsect:stack}).
    \label{fig:boss1244_field}}
\end{figure*}

The BOSS1244 protocluster ($z=2.24\pm0.02$) was discovered via the MAMMOTH (MApping the Most Massive Overdensity Through Hydrogen) technique, which identifies the extreme tail of the matter density distribution at high redshifts, using coherently strong intergalactic \lya absorption in background quasar spectra \citep{Cai:2016gi,Cai:2017ev,Cai:2017df}.
This identification is drawn from the entire SDSS-III quasar spectroscopic sample \citep{Alam:2015go} over a sky coverage of 10,000 deg$^2$, suggesting that BOSS1244 is one of the most massive galaxy protoclusters at cosmic noon.
Deep CFHT broad+narrow band imaging has revealed pronounced overdensities of H$\alpha$ emitters (HAEs), most of which are spectroscopically confirmed by LBT/MMT IR spectroscopy \citep[][also see the left panel of Fig.~\ref{fig:boss1244_field}]{Zheng:2021dw,Shi:2021jt}.
Within a (15\,cMpc)$^{3}$ volume, the density peak region of BOSS1244 in the south west has a galaxy overdensity of $\delta_{\rm gal}=22.9\pm4.9$, where the galaxy overdensity is defined as $\delta_{\rm gal}=\Sigma_{\rm cluster}/\Sigma_{\rm field}-1$ with $\Sigma_{\rm cluster/field}$ corresponding to the HAE number counts per arcmin$^2$ in the protocluster/blank fields at the protocluster redshift.
The derived $\delta_{\rm gal}$ of BOSS1244 is a factor of $\geq$2 larger than that measured in other protoclusters at similar redshifts, \eg, SSA22 at $z=3.1$, which is estimated to have $\delta_{\rm gal}=9.51\pm1.99$ within a (16\,cMpc)$^{3}$ volume \citep{Topping:2018hy}.
This high $\delta_{\rm gal}$ of BOSS1244 corresponds to a present-day total enclosed mass of $(1.6\pm0.2)\times10^{15}$\Msun. BOSS1244 will likely evolve into a galaxy cluster more massive than Coma at $z\sim0$, therefore it is an excellent example of the most extremely overdense environments at $z\geq2$.

We are conducting the \mg slitless spectroscopic survey to obtain followup slitless spectroscopy of three 
  MAMMOTH protoclusters (with BOSS1244 being one of them).
\mg is an \hst cycle-28 medium program (GO-16276, P.I. Wang), awarded 45 primary orbits to acquire deep WFC3/G141-F125W observations in the centers of three of the most massive galaxy protoclusters at $z=2-3$.
The observations in the BOSS1244 field were completed in Feb. 2021, with a total of five different pointings, each at 3-orbit depth.
To mitigate the spectral collision effect of crowded fields, for each pointing, we design a set of 3 individual visits of 1-orbit G141 grism + F125W direct imaging exposures, with an incremental orient offset of 10-15 degrees among the three visits.
This results in $\sim$6 kilo-seconds of G141 spectroscopy (covering 1.1-1.7\,\micron) and $\sim$1.8 kilo-seconds of F125W imaging, useful for the astrometric alignment and wavelength calibration of the paired grism exposure.
In total, \mg acquired 30 kilo-seconds of G141 and 9 kilo-seconds of F125W exposures in the field of the BOSS1244 protocluster (see the right panel of Fig.~\ref{fig:boss1244_field}).

\section{Measurements}\label{sect:measure}

\subsection{Grism redshift and emission line flux}\label{subsect:ELflux}

We utilize the \grzl\footnote{\url{https://github.com/gbrammer/grizli/}} software \citep{Brammer:2021df} to reduce the WFC3/G141 data. Briefly, \grzl analyzes the paired direct imaging and grism exposures in five steps: 
1) pre-processing of the raw WFC3 exposures (e.g. relative/absolute astrometric alignment, flat fielding, variable sky background subtraction due to Helium airglow, etc.),
2) forward modeling full-FoV grism exposures at the visit level,
3) redshift fitting via spectral template synthesis,
4) full-FoV grism model refinement,
5) 1D/2D grism spectrum and emission line map extraction for individual extended objects.

\begin{figure}[!htb]
    \centering
    \includegraphics[width=0.45\textwidth, trim = 0cm 0cm 0cm 0cm, clip]{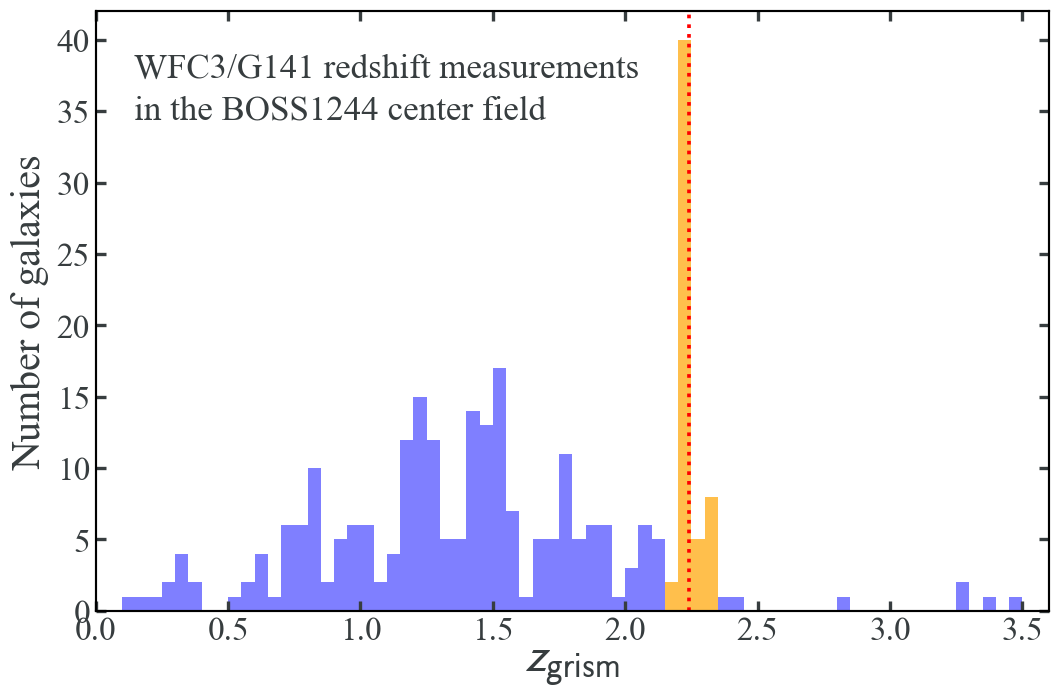}
    \caption{Redshift histograms of galaxies along the line of sight to the BOSS1244 protocluster density peak regions probed by \mg. Here we only include galaxies whose grism redshifts ($z_{\rm grism}$) are considered to be secure according to the joint criteria mentioned in Sect.~\ref{subsect:ELflux}.
    The vertical red dotted line denotes the measured spectroscopic redshift of BOSS1244 as $z_{\rm spec}=2.24\pm0.02$ reported by \citet{Shi:2021jt}.
    Note that $z_{\rm spec}$ was measured with ground-based multi-object spectroscopic instruments with higher wavelength resolution than that of WFC3/G141, so $z_{\rm spec}$ is usually deemed more precise than $z_{\rm grism}$ \citep[see \eg,][]{Matharu:2019ku}.
    The orange histogram marks the redshift range of $z\in[2.15, 2.35]$ where the protocluster member galaxies are selected. The entire line-of-sight structure is mapped out by our deep WFC3/G141 slitless spectroscopy, which independently confirms the extremely overdense nature of the BOSS1244 center field.
    \label{fig:zhist}}
\end{figure}

Our deep grism spectroscopy allows robust redshift measurements following the method detailed in Appendix A of \citet{Wang:2019cf}.
In brief, we fit linear combinations of spectral templates to the observed optimally extracted 1D grism spectra of sources, to search for their best-fit grism redshifts.
The source's grism redshift fit is considered to be secure only if all the following goodness-of-fit criteria are met (\ie the reduced $\chi^2$, the width of the redshift posterior, and the Bayesian information criterion): $\chi^2_{\rm reduced}<2 \wedge \left(\Delta z\right)_{\rm posterior}/(1+z_{\rm peak})<0.005 \wedge \mathrm{BIC}>30$.
As a result, we obtain a sample of 284 galaxies in the redshift range of $z\in[0.15, 4]$, whose grism redshifts are deemed secure according to the above joint selection criteria.
Their redshift histogram is shown in Fig.~\ref{fig:zhist}.
There are in total 55 galaxies (shown in orange in Fig.~\ref{fig:zhist}) with secure grism redshifts of $z\in[2.15, 2.35]$, corresponding to the likely member galaxies of the BOSS1244 protocluster spectroscopically confirmed at $z\sim2.24$ \citep[][marked by the vertical red dotted line in Fig.~\ref{fig:zhist}]{Shi:2021jt}.
Note that the spectroscopic confirmation of BOSS1244's redshift was performed using the ground-based MMT/MMIRS and LBT/LUCI K-band multi-object spectroscopy of HAE candiates, with high wavelength resolution of  $R\sim1200-1900$ \citep{Shi:2021jt}.
Albeit with lower wavelength resolution ($R\sim130$), our grism analysis independently confirms the extremely overdense nature of the BOSS1244 center field, as shown in Fig.~\ref{fig:zhist}.

When deriving the best-fit grism redshifts of our targets of interest, we also fit for their intrinsic nebular emission lines with 1D Gaussian profiles centered at the corresponding wavelengths, and thus obtain their line fluxes.
We thereby obtain flux measurements of multiple emission lines (\OII, \NeIII\footnote{With WFC3/G141 spectral resolution,
$\NeIII\lambda3869$ is blended with H9 on the blue side and \ionp{He}{i}$\lambda3889$+H8 on the red side. We thus exclude \NeIII in our metallicity inference.}, \Hd, \Hg, \Hb, \OIII) in the sample of secure protocluster member galaxies.
To secure accurate metallicity estimates, we select sources with strong detections of \OIII and \OII (both with ${\rm SNR}\gtrsim3$).
This results in a final sample of 39 protocluster member galaxies showing prominent nebular emission features for the subsequent analyses.
Their measured line fluxes are presented in Table~\ref{tab:indvd}.

\subsection{Stellar mass and star-formation rate}\label{subsect:Mstar_SFR}

There has been a wealth of multi-wavelength deep imaging data in the field of BOSS1244.
In addition to the F125W imaging acquired by \mg, 
we take advantage of the archival \hst/WFC3 F160W imaging at 1-orbit depth (obtained by GO-15266, P.I. Cai).
Both the F125W and F160W imaging data have been reduced using the \adriz software,
drizzled onto a common pixel grid with 0\farcs06 pixel scale and \textsc{pixfrac=0.8}, astrometrically aligned to the GAIA DR2 astrometry frame \citep{Brown:2018io}.

We also utilize the existing ground-based imaging data in the field, which includes LBT/LBC Uspec imaging (exposure time: 4.65\,hr, 5$\sigma$ limiting magnitude: 26.67\,mag for a slightly extended source with a 2\arcsec diameter), LBT/LBC $z$-SLOAN imaging (exposure time: 3.92\,hr, 5$\sigma$ limiting magnitude: 25.12\,mag for an extended source with a 2\arcsec diameter), and CFHT $K_{\rm s}$ imaging (exposure time: 7.50\,hr, 5$\sigma$ limiting magnitude: 23.29\,mag for an extended source with a 2\arcsec diameter). The typical seeing condition during the data acquisition of these ground-based data is 0\farcs6-0\farcs9 (D.D. Shi et al. in preparation).

We run \sex in dual mode to perform aperture photometry of the \hst imaging data, with the stacked F125W and F160W mosaics as the detection image to increase the depth.
The total magnitudes in F125W and F160W are then measured from the corresponding isophotal fluxes with proper aperture correction.
To account for the large difference between the \hst and ground-based imaging resolutions, we adopt the \tphot software \citep{Merlin:2016uk} that does template fitting of low-resolution images using a priori knowledge of object position/morphology from high-resolution images.
By minimizing the low-resolution image residuals after removing a combination of models built from high-resolution templates, we obtain accurate photometric measurements from the ground-based imaging data compatible with the isophotal fluxes measured from \hst images \citep{Laidler:2007iy}.
As a consequence, we obtain broad-band photometry over a wide wavelength range of $[1000,7000]$\,\AA\ in the rest frame for the 39 selected protocluster member galaxies at $z\in[2.15, 2.35]$.

To estimate the stellar masses (\Mstar) of our sample galaxies, we use the \bagp software \citep{Carnall:2018gb} to fit the BC03 \citep{Bruzual:2003ck} models of spectral energy distributions (SEDs) to the photometric measurements derived above. We assume the \citet{Chabrier:2003ki} initial mass function, a metallicity range of $Z/Z_{\odot}\in(0, 2.5)$, the \citet{Calzetti:2000iy} extinction law with $A_v$ in the range of (0, 3) and an exponentially declining star formation history with $\tau$ in the range of (0.01, 10)\,Gyr during the SED fitting. The redshifts of our galaxies are fixed to their best-fit grism values, with a conservative uncertainty of $\Delta z/(1+z)\approx 0.003$, following \citet{Momcheva:2016fr}.
Note that we also obtain the actual measurements of $\Delta z/(1+z)$ for our galaxies, where $\Delta z$ is given by the half width between the 16th and 84th percentiles of the redshift posteriors calculated by \grzl.
The median, minimum, and maximum values of $\Delta z/(1+z)$ for our sample is $5.5\times10^{-4}$, $1.3\times10^{-4}$, $2.0\times10^{-3}$, respectively. This is in fair agreement with the finding by \citet{Momcheva:2016fr}. 
Our smaller redshift errors are primarily driven by the fact that our sample sits on a sweet spot of $z\sim2.2-2.3$ where G141 is particularly efficient and accurate at measuring redshifts of prominent line emitters whose multiple emission features (\ie \OIII, \Hb, \OII) are covered simultaneously by this grism element.
For simplicity, we adopt $\Delta z/(1+z)=3\times10^{-3}$ throughout the entire sample as a conservative uncertainty on galaxy redshift in SED fitting.
The nebular emission component is also added into the SED during the fit, since our galaxies are exclusively strong line emitters by selection.

We use the Balmer line luminosities to estimate the instantaneous star-formation rate (SFR) of our sample galaxies. This approach provides a valuable proxy of the ongoing star-formation on a time scale of $\sim$10\,Myr, highly relevant for galaxies displaying strong nebular emission lines.
From our forward-modeling Bayesian method described in Sect.~\ref{subsect:oh12}, we can put stringent constraints on the de-reddened \Hb flux.
Assuming the \citet{Kennicutt:1998ki} calibration and the Balmer decrement ratio of $\Ha/\Hb=2.86$ from the case B recombination for typical \HII regions, we calculate
\begin{align}
    \mathrm{SFR} = 1.6\times10^{-42} \frac{L(\Hb)}{\mathrm{[erg\,s^{-1}]}}\,[\Msun\,\mathrm{yr}^{-1}]
\end{align}
suitable for the \citet{Chabrier:2003ki} initial mass function.
The derived \Mstar and SFR values for our entire galaxy sample are given in Table~\ref{tab:indvd}.

\begin{figure*}[!htb]
    \centering
    \includegraphics[width=0.49\textwidth, trim = 0cm 0cm 0cm 0cm, clip]{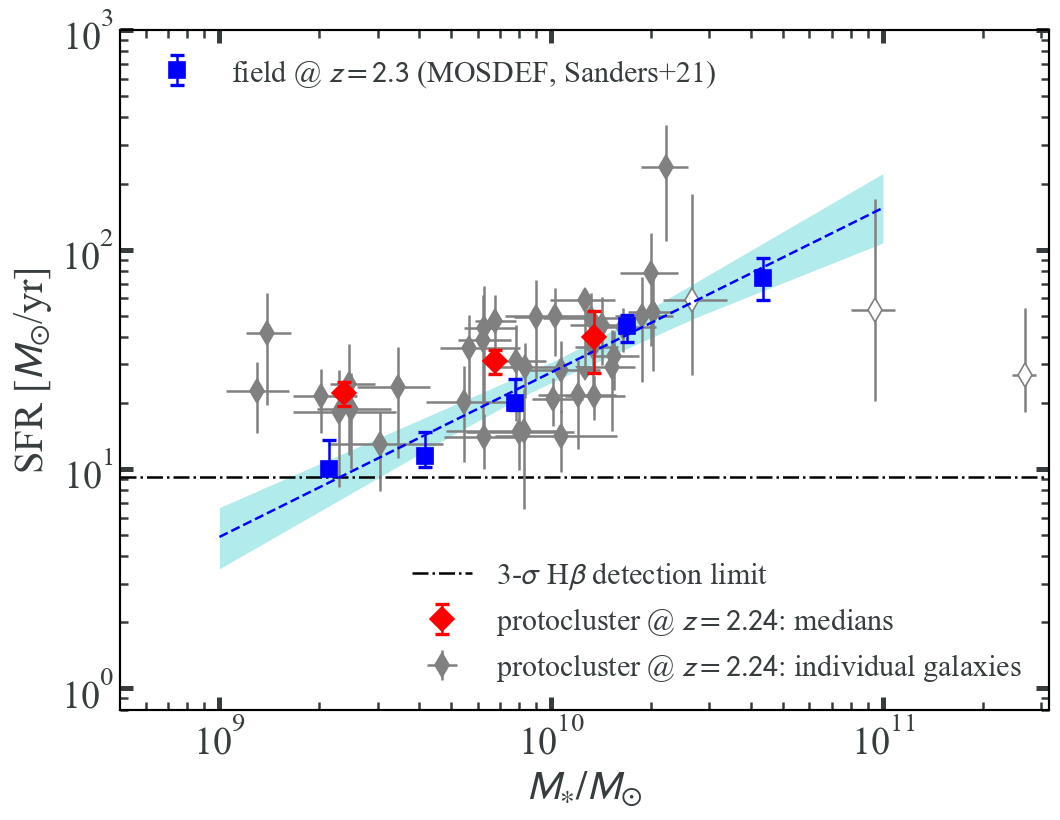}
    \includegraphics[width=0.49\textwidth, trim = 0cm 0cm 0cm 0cm, clip]{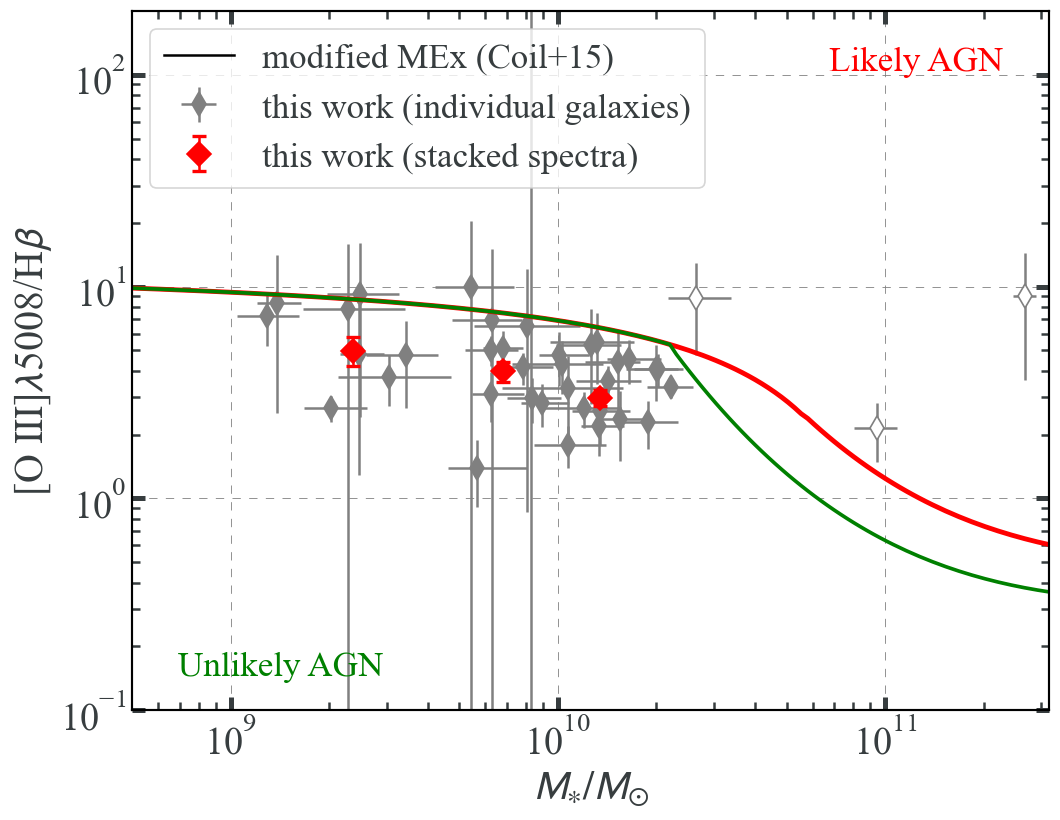}
    \caption{Sample properties of the star-forming member galaxies of the BOSS1244 protocluster at $z\approx2.24$, analyzed in this work.
    {\bf Left}: the SFR-\Mstar relation for our galaxy sample, where the individual measurements are in grey and the median results within the three mass bins are in red.
    The 3$\sigma$ \Hb detection threshold of our grism survey is marked by the horizontal dot-dashed line.
    As a comparison, we also show the loci of similarly selected emission line galaxies in the blank field from the MOSDEF survey at $z\sim2.3$ \citep{Sanders:2021ga}.
    The dashed line and shaded band denote the best-fit power law and 1-$\sigma$ uncertainty intervals of the MOSDEF SFR-\Mstar relation.
    Our median values lie 0.37, 0.18, and 0.06 dex above the MOSDEF relation for the galaxies in low, intermediate, and high mass bin, respectively.
    As seen in \citet{Koyama:2013df}, we observe elevated SFRs in overdense environments compared with those in the field.
    {\bf Right}: the mass-excitation diagram for our galaxy sample. The majority of our galaxies have negligible contamination from AGN ionization given the modified demarcation scheme by \cite{Coil:2015dp} as an updated version of the \citet{Juneau:2014ca} to account for redshift evolution of \OIII/\Hb. The three galaxies marked by hollow diamonds reside in the AGN loci and are therefore eliminated from the subsequent stacking analysis.
    \label{fig:sfms_mex}}
\end{figure*}

In the left panel of Fig.~\ref{fig:sfms_mex}, we show the loci of our protocluster member galaxies in the SFR-\Mstar diagram. The individual galaxies are represented by grey diamonds, whereas the median values within the three mass bins (defined in Sect.~\ref{subsect:stack}) are shown by red diamonds. To facilitate a comparison with field galaxies in the similar \Mstar range at the same $z$, we rely on the galaxy sample at $z\sim2.3$ from the MOSDEF survey \citep{Kriek:2015ek,2015ApJ...801...88S}, because the MOSDEF galaxies are also emission-line selected with SFRs estimated from Balmer line fluxes.
The best-fit SFR-\Mstar relation for the $z\sim2.3$ MOSDEF galaxies is shown as the blue dashed line in the left panel of Fig.~\ref{fig:sfms_mex}, with the cyan shaded band representing its 1-$\sigma$ uncertainty.
\citet{Sanders:2021ga} show that the MOSDEF sample is representative of galaxies on the star-forming main sequence (SFMS) at $z\sim2.3$ in the stellar mass range of $10^9-10^{10.75}$\Mstar.
The median values from our protocluster galaxy sample show enhanced SFRs by 0.06, 0.18, and 0.37 dex as compared to the MOSDEF SFR-\Mstar relation.
To make a fair comparison, we also need to account for the different line flux limits from the two surveys.
The 3$\sigma$ limiting line flux for our 3-orbit G141 \mg exposures is $1.7\times10^{-17}$~\Funit \citep[also see][]{Atek:2010gc,Momcheva:2016fr}.
This is 0.28\,dex higher than the MOSDEF detection limit, which is $1.5\times10^{-17}$~\Funit at 5$\sigma$ within a 2\,hr observation reported by \citet{Kriek:2015ek}.
Our \Hb line flux detection threshold thus translates into an SFR limit of 9.2\,\Msun\,yr$^{-1}$, represented by the dot-dashed line in Fig.~\ref{fig:sfms_mex}.
Comparing our SFR threshold with the 1-$\sigma$ spread of the MOSDEF galaxies,
we see that our detection threshold allows complete sampling of the SFMS at $z\sim2.24$ down to $\Mstar\sim10^{9.7}\Msun$.
We therefore argue that the SFR increases detected in our intermediate and high mass bins are robust and primarily stem from the environmental effect.
We also tentatively observe that the star formation enhancement induced by overdense environments is more prominent for lower mass galaxies: $\Delta$SFR$\approx$+0.18 dex in our galaxies with $\log(\Mstar/\Msun)\in[9.7, 10.0)$ and $\Delta$SFR$\approx$+0.06 dex in our galaxies with $\log(\Mstar/\Msun)\in[10.0, 10.4)$. 
\citet{Koyama:2013df} also found similarly small SFR offset between the cluster and field environments from galaxies with $\log(\Mstar/\Msun)\in[10.0, 11.4)$ at $z\sim2.2$.
The increase in SFR from our sample is likely due to the extremely overdense nature of the BOSS1244 protocluster, which boosts star formation in the member galaxies, through continued gas supply replenished by the cold-mode accretion from the cosmic web \citep{Dekel:2009fz,2009MNRAS.395..160K} and/or by the recycling of metal pre-enriched gas blown out by galactic feedback \citep{Oppenheimer:2008bu}.
This environmental effect is also stronger in lower mass galaxies since more massive galaxies intrinsically live in denser environment.

\subsection{AGN contamination}\label{subsect:AGN}

Before carrying out the stacking analysis and metallicity inference of our sample galaxies, we make sure that the individual galaxies selected for stacking are primarily photoionized by massive stars in the \HII regions, rather than by active galactic nuclei (AGNs). The galaxies in our sample have measured \OIII and \Hb fluxes from grism spectroscopy, and \Ha from ground-based LBT/MMT IR spectroscopy, yet not always \NII fluxes due to its intrinsic faintness, which precludes us from using the BPT diagram \citep{Baldwin:1981ev} to separate the loci of AGNs and star-forming galaxies.
\citet{Juneau:2014ca} proposed an effective approach coined the mass-excitation (MEx) diagram, using \Mstar as a proxy for \NII/\Ha, which functions well in differentiating \HII regions from AGNs in emission line galaxies at $z\sim0$ from SDSS DR7.
\citet{Coil:2015dp} further modified the MEx demarcation by shifting the separation curves to the high-\Mstar end by 0.75\,dex, which is shown to be more applicable to the MOSDEF sample at $z\sim2.3$.
We thus rely on this modified version of MEx to prune AGN contamination from our galaxy sample.
As shown in the right panel of Fig.~\ref{fig:sfms_mex}, there are three galaxies that pass the previous selection criteria (secure grism redshift at $z\in[2.15,2.35]$, \OIII and \OII having $SNR\gtrsim$3) that are categorised as AGNs, which are hence removed from the stacking analysis  below.

\subsection{Spectral stacking}\label{subsect:stack}

From the previous procedures, we secured in total 36 spectroscopically confirmed member galaxies of the BOSS1244 massive protocluster, which are undergoing active star formation.
We first divide our entire galaxy sample into three mass bins: 
$\log(\Mstar/\Msun) \in$~[10.0, 10.4) (high-mass bin), [9.7, 10.0) (intermediate-mass bin), [9.0, 9.7) (low-mass bin).
There are 17, 11, and 8 galaxies in the high-, intermediate- and low-mass bins, respectively.

Then we adopt the following stacking procedures, similar to those utilized in \citet{Henry:2021ju}.
\begin{enumerate}
    \item Subtract continuum models from the observed grism spectra, combined from multiple orients. The continuum models are constructed using the \grzl software.
    \item Normalize the continuum-subtracted spectrum of each object using their measured \OIII flux, to avoid excessive weighting towards objects with stronger line fluxes.
    \item De-redshift each normalized spectrum to its rest frame on a common wavelength grid.
    \item Take the median value of the normalized fluxes at each wavelength grid.
    \item Re-create the stacked spectra 1000 times with bootstrapping replacement, and adopt the standard deviation as the measurement uncertainty.
\end{enumerate}

\begin{figure*}[!htb]
    \centering
    \includegraphics[width=.8\textwidth, trim = 0cm 0cm 0cm 0cm, clip]{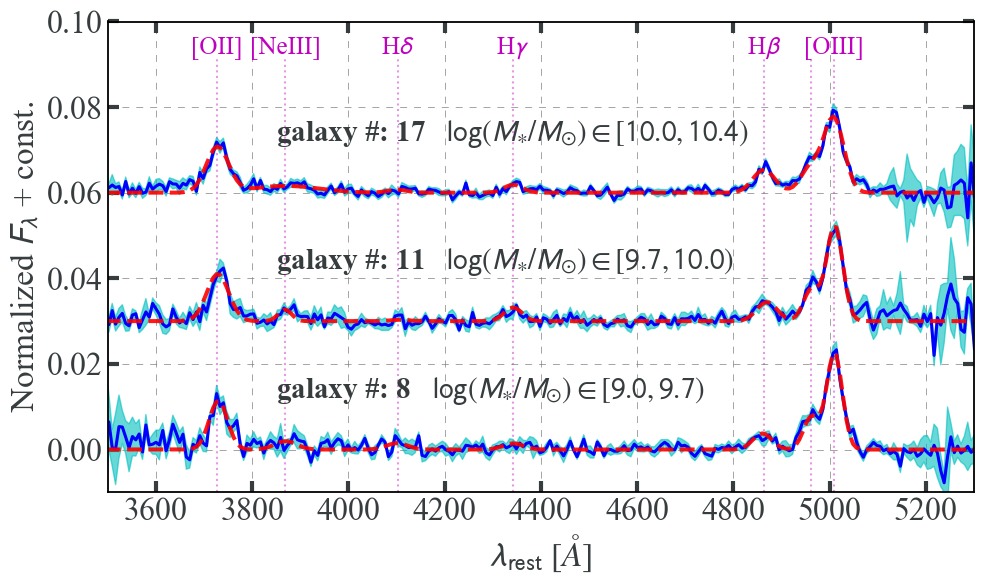}
    \caption{Stacked grism spectra for galaxies residing in the three mass bins.
    For the high-, intermediate-, and low-mass bins, we secured a number of 17, 11, and 8 galaxies at $z\approx2.24$, respectively, with the corresponding mass range highlighted above the spectra.
    In each set of spectra, the blue curves represent the median stacked spectrum, the cyan bands mark the bootstrapped flux uncertainties, and the red dashed curves show the best-fit Gaussian fits to multiple emission lines (\ie the \OIII$\lambda\lambda$4960,5008 doublets, \Hb, \Hg, \Hd, \NeIII, and \OII).
    The details of the stacking procedures are presented in Sect.~\ref{subsect:stack}.
    \label{fig:stacked_spec}}
\end{figure*}

Next we fit multiple Gaussian profiles to the stacked spectra to derive line flux ratios within each mass bin.
The multiple Gaussian profiles are centered at the corresponding rest-frame wavelengths of emission lines of \OII, \NeIII, \Hd, \Hg, \Hb and the \OIII$\lambda\lambda$4960,5008 doublets (whose amplitude ratio is fixed to 1:2.98 following \citet{Storey:2000jd}).
We use the LMFIT software \footnote{\url{https://lmfit.github.io/lmfit-py/}} to perform the non-linear least square minimization\footnote{Albeit not utilized here, the \grzl software package also includes some functionalities for least square minimization.}, yielding a $\chi^2_{\rm reduced}\sim1-1.5$ for the fits.
The resultant stacked grism spectra for the three mass bins are shown in Fig.~\ref{fig:stacked_spec} with the best-fit emission line models marked by the red dashed curves.
The measured quantities within each mass bin are summarized in Table~\ref{tab:stack}.
Interestingly, we see a clear increase in \OIII/\Hb as the median \Mstar decreases, reflecting the MEx demarcation scheme shown in Fig.~\ref{fig:sfms_mex}.

\subsection{Gas-phase metallicity}\label{subsect:oh12}

We follow our previous series of work \citep{Wang:2016um,Wang:2019cf,Wang:2020bp} to constrain jointly metallicity (\oh), nebular dust extinction ($A_v$), and de-reddened \Hb flux ($f_{\Hb}$), using our forward modeling Bayesian inference method.
This method is superior to the conventional way of first calculating line flux ratios and then converting the ratios to metallicity, because it can marginalize faint lines (e.g. \Hb) with low SNRs, as demonstrated in Fig.~6 of \citet{Wang:2020bp}. The likelihood is defined as 
\begin{align}\label{eq:chi2}
    \mathrm{L}\propto\exp\left(-\frac{1}{2}\cdot\sum_i \frac{\(f_{\el{i}} - R_i \cdot 
    f_{\Hb}\)^2}{\(\sigma_{\el{i}}\)^2 + \(f_{\Hb}\)^2\cdot\(\sigma_{R_i}\)^2}\right).
\end{align}
where $f_{\el{i}}$ and $\sigma_{\el{i}}$ represent the emission line (\eg \OII, \Hg, \Hb, \OIII) flux and its uncertainty, corrected for dust attenuation using the \citet{Calzetti:2000iy} extinction law with $A_v$ drawn from the parameter sampling.
In practice, this extinction correction is performed on-the-fly with the parameter sampling process, since each sampled value of $A_v$ is given by each random draw in the three-dimensional parameter space spanned by (\oh, $A_v$, $f_{\Hb}$) from the likelihood using the Markov Chain Monte Carlo (MCMC) algorithm.
$R_i$ refers to the expected line flux ratios, being either the Balmer decrement of $\Hg/\Hb=0.47$ or the metallicity diagnostics of $\OIII/\Hb$ and $\OII/\Hb$, with $\sigma_{R_i}$ being their intrinsic scatters.
There exists a wide range of approaches in converting the strong line diagnostics to metallicities \citep[see Appendix C in][for a summary]{Wang:2019cf}. Different choices of metallicity calibrations can result in normalization offsets as high as 0.7 dex \citep[see \eg][]{Kewley:2008be}.
In this work we adopt the purely empirical strong line calibrations prescribed by \citet[][hereafter B18]{Bian:2018km}, based on a sample of local analogs of high-$z$ galaxies according to the location on the BPT diagram.
The coefficients for these strong line calibrations are given in Table~\ref{tab:coef}.
These calibrations are shown to reproduce well the observed ratios of \OIII/\Hb and \OII/\Hb for a sample of 18 galaxies at $z\sim2.2$ whose metallicities are derived using the direct electron temperature method \citep{Sanders:2021ga}.
Furthermore, the B18 calibrations are also employed in the metallicity measurements for the MOSDEF galaxies, so that it is straightforward to derive the metallicity offset between galaxies in cluster and field environments.
We thus refer to our metallicity estimates obtained with the B18 calibrations as our default results.
In addition, we also provide our metallicity estimates assuming the \citet[][hereafter C17]{Curti:2017fn} strong line calibrations, whose coefficients are also given in Table~\ref{tab:coef}.
An advantage of using oxygen lines rather than nitrogen lines in inferring metallicity is that the result is free from the systematics of using locally calibrated \NII/\Ha as a high-$z$ metallicity diagnostic, given the locus offset of star-forming galaxies at $z\sim0$ and $z\gtrsim2$ in the BPT diagram, whose origin is still under debate \citep{2014ApJ...795..165S,2015ApJ...801...88S}.
The \emc software \citep{ForemanMackey:2013io} is employed to perform the MCMC parameter sampling.
As a result, we have obtained the key physical properties (\oh, \Mstar, and SFR) of our sample galaxies as well as the stacked spectra within three mass bins.
The measured properties for the individual galaxies and the stacked mass bins are shown in Tables~\ref{tab:indvd} and \ref{tab:stack} respectively.

\begin{deluxetable}{lcccccccccccc}
    \tablecolumns{5}
    \tablewidth{0pt}
    \tablecaption{Coefficients for the emission line flux ratio diagnostics used in this work.}
\tablehead{
    \colhead{$\log(R)$} &
    \colhead{$c_0$} &
    \colhead{$c_1$} &
    \colhead{$c_2$} &
    \colhead{$c_3$} 
}
\startdata
    \multicolumn{5}{c}{Strong line calibrations of \citet[][B18]{Bian:2018km}\tablenotemark{a}}   \\
    \noalign{\smallskip}
    \OIII$\lambda\lambda$4960,5008/\Hb   &   43.9836    &   -21.6211   &   3.4277  &   -0.1747   \\
    \OII/\Hb    &   78.9068    &   -45.2533   &   7.4311  &   -0.3758   \\
    \noalign{\smallskip}\hline\noalign{\smallskip}
    \multicolumn{5}{c}{Strong line calibrations of \citet[][C17]{Curti:2017fn}\tablenotemark{b}}   \\
    \noalign{\smallskip}
    \OIII/\Hb   &   -0.277 &  -3.549 & -3.593 & -0.981  \\
    \OII/\Hb    &   0.418 &  -0.961 & -3.505 & -1.949   \\
    \noalign{\smallskip}\hline\noalign{\smallskip}
    \multicolumn{5}{c}{Balmer decrement}   \\
    \noalign{\smallskip}
    \Hg/\Hb   &   -0.3279    &   \nodata   &   \nodata  &   \nodata \\
\enddata
    \tablecomments{The empirical flux ratios are computed in the polynomial functional form of $\log{R} = \sum_{i} c_i\cdot x^i$, where $x=\oh-8.69$ for the \citet{Curti:2017fn} calibrations and $x=\oh$ for the \citet{Bian:2018km} ones.
    }
    \tablenotetext{a}{We note that the \OIII/\Hb calibration reported in \citet{Bian:2018km} in fact refers to the flux ratio between \OIII$\lambda\lambda$4960,5008 and \Hb, \ie, higher than \OIII/\Hb by a factor of (3.98/2.98).}
    \tablenotetext{b}{The intrinsic scatter of 0.09(0.11) dex for the \OIII/\Hb(\OII/\Hb) calibrations, as reported in \citet{Curti:2017fn}, have been included in the metallicity inference.}
\label{tab:coef}
\end{deluxetable}


\section{The mass-metallicity relation in the BOSS1244 protocluster environment}\label{sect:mzr}

\begin{figure*}[!htb]
    \centering
    \includegraphics[width=0.48\textwidth, trim = 0cm 0cm 0cm 0cm, clip]{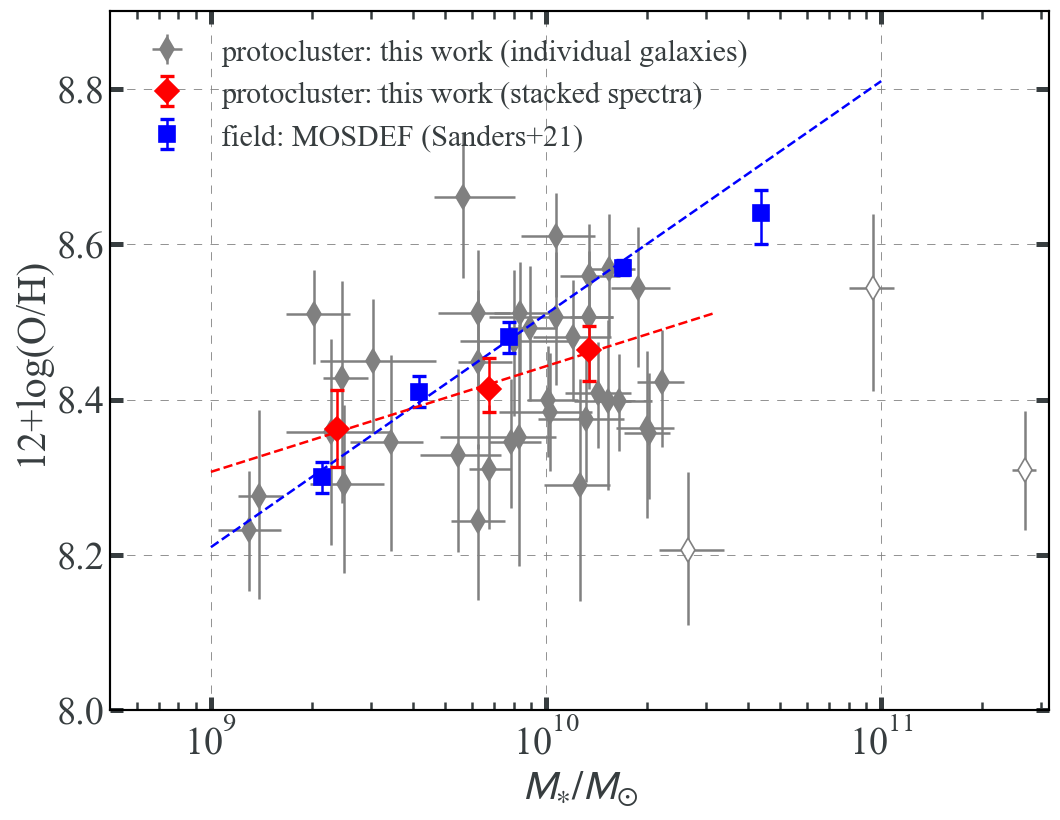}
    \includegraphics[width=0.49\textwidth, trim = 0cm 0cm 0cm 0cm, clip]{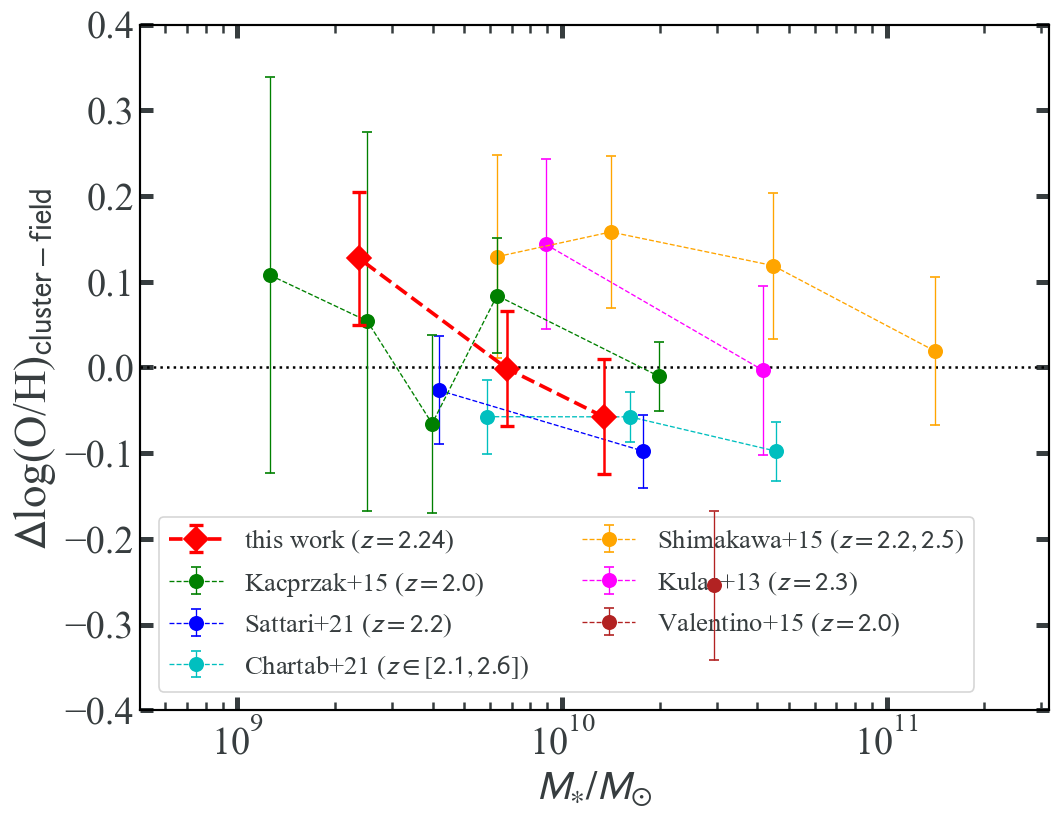}
    \caption{Metallicity measurements for the star-forming member galaxies of the BOSS1244 protocluster at $z\approx2.24$, analyzed in this work.
    {\bf Left}: the mass-metallicity relation (MZR) for our galaxy sample (see Sect.~\ref{subsect:oh12} for our method of inferring metallicities), where the individual measurements are in grey and the stacked results are in red.
    As in Fig.~\ref{fig:sfms_mex}, the hollow diamonds mark the three AGNs categorized according to the modified MEx diagram, and thus do not contribute to the stacked results.
    We also show the MZR measured in field from the MOSDEF survey at $z\sim2.3$ \citep{Sanders:2021ga}.
    Note that all the metallicity measurements shown here are derived using the same set of strong line calibrations by \citet[][B18]{Bian:2018km}, to facilitate a direct comparison.
    We find a significantly shallower MZR slope from our galaxy sample in overdense environments, compared with the field result at similar redshifts.
    {\bf Right}: the metallicity offset between protocluster and field galaxies at $z\geq2$, as a function of stellar mass. 
    To compensate for the SFR differences and isolate the environmental effect for our protocluster member galaxies,
    we compute their reference values of field metallicity according to the fundamental metallicity relation established in \citet{Sanders:2021ga} given the measurements of \Mstar and SFR for our sample galaxies (see Sect.~\ref{sect:mzr} for more details).
    Our results show an enhancement of metallicity in low-mass galaxies and a slight deficiency of metallicity in high-mass galaxies residing in overdense environments, which is in accordance with the literature results.
    The detailed properties of these high-$z$ overdense environments where the literature results on metallicity offsets have been made are presented in Table~\ref{tab:clusters}.
    \label{fig:mzr}}
\end{figure*}

In this section, we present our key result, the measurement of gas-phase metallicities as a function of galaxy stellar mass in the density peak region of the massive protocluster BOSS1244 at $z\sim2.24$.
This is to our knowledge the first ever effort to derive the MZR at $z\gtrsim2$ in overdense environments using grism slitless spectroscopy. The high throughput and sensitivity of the WFC3/G141 channel, coupled with the relatively low IR sky background in space, allow us to measure the MZR down to \Mstar as low as $\sim10^9\,\Msun$ at $z\approx2.24$.
The left panel of Fig.~\ref{fig:mzr} shows the loci of our protocluster member galaxies in the \Mstar versus \oh diagram, again with individual sources marked in gray, while the stacked results are in red.
We perform linear regression to the stacks to derive the following MZR in the \Mstar range of $[10^{9}, 10^{10.4}]$\,\Msun: $\oh^{\rm B18} = \left(0.136\pm0.018\right) \times \log(\Mstar/\Msun) + \left(7.082 \pm 0.175 \right)$, , shown as the red dashed line.
We stress that these measurements are obtained assuming the \citet[][B18]{Bian:2018km} strong line calibrations, for the sake of a straightforward comparison with the MOSDEF metallicity measurements, which are derived under the same set of calibrations.
In Tables~\ref{tab:indvd} and \ref{tab:stack}, we also provide our metallicity estimates for individual galaxies and stacks, using the C17 calibrations instead.
The C17 MZR for our sample galaxies reads $\oh^{\rm C17} = \left(0.106\pm0.013\right) \times \log(\Mstar/\Msun) + \left(7.329 \pm 0.132 \right)$, showing similar values of slopes and intercepts with our default results.

{
\tabletypesize{\scriptsize}
\tabcolsep=2pt
\begin{deluxetable*}{lcccccccccccc}
    \tablecolumns{13}
    \tablewidth{0pt}
    \tablecaption{Measured properties of the stacked spectra.}
\tablehead{
    \colhead{mass bin} &
    \colhead{$N_{\rm gal}$} &
    \colhead{log($M_{\ast}/M_{\odot}$)} &
    \colhead{$M_{\ast}^{\rm med}$\tablenotemark{a}}  &
    \colhead{SFR}   &
    \colhead{\OIII/\Hb}  &
    \colhead{\OII/\Hb}  &
    \colhead{\Hg/\Hb}  &
    \colhead{\OIII/\OII}  &
    \colhead{$f_{\OIII}$}  &
    \colhead{\oh\tablenotemark{b}}  &
    \colhead{$\Delta~\mathrm{log(O/H)_{cluster-field}}$\tablenotemark{c}}  &
    \colhead{\oh\tablenotemark{d}}  \\
    & & range & [$M_{\odot}$] & [$M_{\odot}$/yr] &  &  &  &  & [$10^{-17}$\Funit]  &  \multicolumn{2}{c}{using B18 calibrations}  &  using C17 calibrations
}
\startdata
    high    &  17   & [10.0, 10.4) & 10$^{10.13}$ & $40.0\pm12.7$ & $2.99\pm0.26$ & $2.02\pm0.28$ & $0.30\pm0.11$ & $1.48\pm0.16$ & $9.82\pm1.68$  & $8.46_{-0.04}^{+0.03}$  & $-0.06\pm0.07$ &  $8.40_{-0.06}^{+0.04}$  \\
    intermediate & 11& [9.7, 10.0) &  10$^{9.83}$ & $30.9\pm3.8$  & $3.97\pm0.43$ & $2.66\pm0.40$ & $0.62\pm0.13$ & $1.49\pm0.16$ & $9.89\pm1.52$  & $8.41_{-0.03}^{+0.04}$  & $0.00\pm0.07$  &  $8.36_{-0.06}^{+0.05}$  \\
    low    &  8       & [9.0, 9.7) &  10$^{9.38}$ & $22.1\pm2.8$  & $4.98\pm0.78$ & $2.93\pm0.54$ & $0.41\pm0.22$ & $1.69\pm0.19$ & $7.37\pm1.52$  & $8.36_{-0.05}^{+0.05}$  & $0.13\pm0.08$  &  $8.32_{-0.07}^{+0.06}$ 
\enddata
    \tablecomments{The multiple emission line flux ratios are measured from the stacked spectra shown in Fig.~\ref{fig:stacked_spec}. The $f_{\OIII}$ and SFR results refer to the median value of galaxies within each mass bin, with 1-$\sigma$ uncertainty represented by the standard deviation.
    }
    \tablenotetext{a}{The median stellar mass of galaxies within each mass bin.}
    \tablenotetext{b}{The metallicity inference derived from the measured line flux ratios in the stacked spectra presented in each corresponding row, using the method described in Sect.~\ref{subsect:oh12}.
    Here we use the strong line calibrations prescribed by \citet[][B18]{Bian:2018km} as our default results. See Table~\ref{tab:coef} for the relevant coefficients.}
    \tablenotetext{c}{The difference in metallicity between our galaxies in overdense environments and the field measurements inferred from the fundamental metallicity relation prescribed by \citet{Sanders:2021ga}.
    The intrinsic scatter of 0.06 dex has been combined in quadrature into the measurement uncertainties.}
    \tablenotetext{d}{The metallicity inference derived using the strong line calibrations given by \citet{Curti:2017fn}. All other assumptions and data are the same as our default results using the B18 calibrations.}
\label{tab:stack}
\end{deluxetable*}
}


As a comparison, we plot the MZR measured in blank fields at the similar redshift from the MOSDEF sample \citep{Sanders:2021ga}, using the same set of metallicity diagnostics as our default results. 
The best-fit single power law form of the MOSDEF MZR at $z\sim2.3$ is $\oh_{\rm field,Sanders+21}^{\rm B18} = \left(0.30\pm0.02\right) \times \log(\Mstar/\Msun) + \left(5.51 \pm 0.20 \right)$ over the \Mstar range of $[10^{9}, 10^{10.75}]$\,\Msun, represented by the blue dashed line.
We also compare with the most up-to-date grism MZR analysis in blank fields by \citet{Henry:2021ju} in the redshift range of $z\sim1-2$. By stacking a large number of galaxies, those authors are able to extend the metallicity determinations to very low \Mstar.
Here we focus on the \Mstar range in overlap with our protocluster MZR measurement and derive a best-fit single power-law relation of $\oh_{\rm field,Henry+21}^{\rm C17} = \left(0.180\pm0.007\right) \times \log(\Mstar/\Msun) + \left(6.619 \pm 0.066 \right)$.
Note that the metallicities presented in \citet{Henry:2021ju} are derived assuming the C17 calibrations.
In both cases, the slope of the field MZR is significantly steeper than that of our MZR measured in arguably the most extremely overdense environment at cosmic noon.

As discussed in Sect.~\ref{subsect:Mstar_SFR}, we are aware of the SFR differences between our stacks and those of the MOSDEF sample at the same redshift.
There is growing consensus about the existence of a redshift-invariant three-parameter relation among {\Mstar, SFR, \oh} --- termed the fundamental metallicity relation (FMR) --- that governs the behaviors of the MZR,
such that at fixed stellar mass, galaxies with higher SFRs display lower metallicities \citep{Mannucci:2009et,Sanders:2017vi,Curti:2019dm}.
Since our galaxies in the low mass bin ($\Mstar/\Msun\in[10^9,10^{9.7}]$) lie $\sim$0.37 dex above the SFMS on average, it is expected that they should be more metal poor by $\sim$0.07 dex than the MOSDEF stack of the same mass, following $\mathrm{O/H}\propto \mathrm{SFR}^{-0.19\pm0.04}$ \citep{Sanders:2021ga}.
However, the metallicity in our low mass bin is estimated to be more metal rich by $\sim$0.06 dex, due to the fact that our galaxies reside in overdense environment.
To isolate this environmental effect, we calculate the reference value of field metallicity for our galaxy sample using the FMR prescribed by \citet[][\ie their Eq.~10]{Sanders:2021ga} to derive the metallicity offsets from our stacked measurements in the BOSS1244 protocluster core.
The resultant metallicity differences between the cluster and field environments after correcting for the SFR differences are shown in the right panel of Fig.~\ref{fig:mzr}.
We found that our galaxies in the low-mass bin ($\Mstar/\Msun\in[10^9,10^{9.7}]$) are on average metal enriched by $0.13\pm0.08$\,dex than the coeval galaxies in the field of the same mass, whereas the galaxy sample in our high-mass bin ($\Mstar/\Msun\in[10^{10},10^{10.4}]$) shows a very small metal deficiency compared with their field reference value calculated from the FMR, albeit not significant given the measurement uncertainty ($-0.06\pm0.07$\,dex).

We also collect a large sample of $z\gtrsim2$ measurements of metallicity offset between cluster and field from the literature.
The detailed properties of these overdense environments where the metallicity offsets have been measured are summarized in Table~\ref{tab:clusters}.
\citet{Sattari:2021ur} analyzed a sample of 19 member galaxies in a spectroscopically confirmed protocluster CC2.2 at $z=2.23$ \citep{Darvish:2020cy} and found that CC2.2 member galaxies with $\Mstar/\Msun\in[10^{9.9},10^{10.9}]$ are metal deficient by $0.10\pm0.04$ dex.
Similarly, \citet{Valentino:2015gv} detected a 0.25\,dex metal deficiency in a sample of 6 galaxies in the mass range of $\Mstar/\Msun\in[10^{10},10^{11}]$ in an X-ray-selected cluster CL J1449+0856 spectroscopically confirmed at $z=2$.
In contrast, other works show the opposite trend: galaxies residing in overdensities with $\Mstar\gtrsim10^{10}\,\Msun$ are more metal rich by as much as 0.16 dex than their field counterparts \citep[see \eg,][]{Kulas:2013ica,Shimakawa:2015iwa}.
The only paper pre-dating our work that attempts to probe the metallicity offset down to $10^9$\,\Msun suffers from low SNR \citep{Kacprzak:2015cg}.
Using grism spectroscopy, we for the first time measure the environmental effect in metal enrichment at cosmic noon down to $\sim2\times10^9$\,\Msun, detecting a metallicity increase in protocluster member galaxies at 1.6$\sigma$ significance.
In Table~\ref{tab:clusters}, we also list the cluster-centric radius range for each protocluster member galaxy sample, normalized by the virial radius of the corresponding protocluster.
The virial radius can be estimated by $R^{\rm cl}_{200} = G M^{\rm cl}_{200}/\left(3\sigma_{\rm los}^2\right)$ \citep{Carlberg:1997dq}, with the caveat that it likely underestimates the size of the cluster-centric region if the system is far from virialization --- rightfully so for \emph{proto}clusters.
Here $M^{\rm cl}_{200}$ and $\sigma_{\rm los}$ represent the dynamical mass and the line-of-sight velocity dispersion of the protocluster.
We see that the cluster-centric radius range for our galaxies is in general agreement with those
where the literature metallicity offset analyses were undertaken.
This verification is important since the physical conditions (\eg merger rates and gas supply through the cosmic web) will likely be different for regions close to or afar from the cores of clusters, which can affect the measurements of the MZR \citep{Shimakawa:2018ft}.

{
\tabletypesize{\scriptsize}
\tabcolsep=2pt
\begin{deluxetable*}{lcccccccccccccc}
    \tablecolumns{10}
    \tablewidth{0pt}
    \tablecaption{Basic properties of the $z\geq2$ overdense environments where the metallicity offsets between protocluster and field galaxies have been measured in the literature.}
\tablehead{
    \colhead{Reference} &
    \colhead{Instrument\tablenotemark{a}} &
    \multicolumn{6}{c}{Galaxy Protoclusters}  &
    \multicolumn{3}{c}{Protocluster member galaxies}  \\
     & &
    \multicolumn{6}{c}{\hrulefill}  &
    \multicolumn{3}{c}{\hrulefill}  \\
     & &
    \colhead{Name} &
    \colhead{$z$} &
    \colhead{$\delta_{\rm gal}$\tablenotemark{b}} &
    \colhead{$M^{\rm cl}_{z=0}$\tablenotemark{c}} &
    \colhead{$M^{\rm cl}_{200}$\tablenotemark{d}} &
    \colhead{$R^{\rm cl}_{200}$\tablenotemark{e}} &
    \colhead{Number counts} &
    \colhead{\Mstar range} &
    \colhead{Cluster-centric radius}    \\
     & & & & & [10$^{15}$\Msun] & [10$^{14}$\Msun] & [Mpc] & & [$\log(\Mstar/\Msun)$] & [$r/R^{\rm cl}_{200}$] 
}
\startdata
    \citet{Kulas:2013ica}  & Keck/MOSFIRE  & HS 1700+643    & 2.3  & 6.9    & 1.2   & \nodata   & \nodata  & 23    & [10.0, 10.6]  &  \nodata   \\
    \citet{Kacprzak:2015cg}  & Keck/MOSFIRE  & \nodata    & 2.095  & \nodata    & \nodata  & 0.94  & 0.44   & 43    & [9.1,10.3]  & <6.8    \\
    \multirow{2}{*}{\citet{Shimakawa:2015iwa}} & \multirow{2}{*}{Subaru/MOIRCS} & PKS 1138−262  & 2.156     & \nodata   & \nodata   & 1.71  & 0.53  & 27  & [9.8,11.15]   & <4.5   \\
     & &    USS 1558−003    &  2.533    & \nodata   & \nodata   & 0.87  & 0.38  & 36  & [9.8,11.15]   & <4.5 \\
    \citet{Valentino:2015gv}    & Subaru/MOIRCS     & CL J1449+0856     & 1.99  & \nodata   & \nodata   & 0.53  & 0.40  & 6     & [10.0, 11.0]  & <2.0   \\
    \citet{Chartab:2021ed}  & Keck/MOSFIRE  & \nodata   & [2.1, 2.6]\tablenotemark{f}    & 1.48\tablenotemark{f}   & \nodata   & \nodata   & \nodata    & 110  & [9.8, 10.7] & \nodata   \\
    \citet{Sattari:2021ur}  & Keck/MOSFIRE  & CC2.2     &   2.23    & 6.6   &   0.9     & 1.4   & 0.49  & 19    & [9.6,10.2]    & <4.9   \\
    This work   & \hst/WFC3     & BOSS1244  & 2.24  &   22.9$\pm$4.9    & 1.6$\pm$0.2  & 0.3$\pm$0.5    & 0.30$\pm$0.15     & 36    & [9.0,10.4]    & <6.0
\enddata
    \tablenotetext{a}{The spectroscopic instruments used in the respective works to obtain metallicity measurements. Our work presents the first ever effort using space-based grism spectroscopy afforded by \hst/WFC3.}
    \tablenotetext{b}{The galaxy overdensity $\delta_{\rm gal}=\Sigma_{\rm cluster}/\Sigma_{\rm field}-1$ where $\Sigma_{\rm cluster/field}$ correspond to the galaxy surface densities measured in the overdense/blank fields.}
    \tablenotetext{c}{The present-day total enclosed mass derived from $\delta_{\rm gal}$.}
    \tablenotetext{d}{The dynamical mass at the epoch of observation derived from the line-of-sight velocity dispersion measured in member galaxies. Note that $M^{\rm cl}_{200}$ underestimates the total mass if the system is far from virialization.}
    \tablenotetext{e}{The virial radius of galaxy cluster given by its dynamical mass and line-of-sight velocity dispersion, \ie, $R^{\rm cl}_{200} = G M^{\rm cl}_{200}/\left(3\sigma_{\rm los}^2\right)$.}
    \tablenotetext{f}{The overdense environments analyzed in this work are identified from the CANDELS fields across a wide redshift range. The quoted value of $\delta_{\rm gal}$ here is the average density contrast measured in their overdense galaxy sample.}
\label{tab:clusters}
\end{deluxetable*}
}


A possible explanation for this metallicity enhancement observed in our low-mass galaxies is efficient recycling of wind material ejected by galactic feedback.
Using cosmological hydrodynamic simulations, \citet{Oppenheimer:2008bu} found that the environmental density is the primary factor that controls the recycling of gas and metals ejected by momentum-driven winds: a denser environment slows the wind particles, allowing them to be re-accreted on a shorter time scale. 
As our low-mass galaxies (with $\Mstar/\Msun\in[10^9,10^{9.7}]$) reside in a more overdense environment compared with field galaxies of the same mass, they are more capable of retaining their stellar nucleosynthesis yields.
Meanwhile, once this ``galactic fountain'' material gets re-accreted onto galaxy disks, it provides metal pre-enriched gas as fuel for star formation, and elevates the SFR as observed in Fig.~\ref{fig:sfms_mex}.

On the other hand, more massive galaxies are intrinsically less prone to the ejective feedback mechanism due to deeper potential wells \citep{ElBadry:2017ge}.
This explains why galaxies in our higher mass bins do not show increased metallicities.
Instead, we observe a slight metal deficiency together with a $\sim$0.3\,dex increase in SFR from our high-mass galaxy sample.
This is likely ascribed to cold-mode gas accretion \citep{Dekel:2009bn}, driven by the massive dark matter halo associated with the BOSS1244 protocluster.
The southwest region of BOSS1244 (targeted by \mg and shown in Fig.~\ref{fig:boss1244_field}) has a galaxy overdensity of $\delta_{\rm gal}=22.9\pm4.9$, and is predicted to become a Coma-type cluster at $z\sim0$ with a total enclosed mass of $(1.6\pm0.2)\times10^{15}\Msun$. However at the time of observation its measured line-of-sight velocity dispersion is $\sigma_{\rm los}=405\pm202$\,km\,s$^{-1}$, translated into a dynamical mass of $M_{200}=(3.0\pm5.0)\times10^{13}$\,\Msun \citep{Shi:2021jt}.
This means that the BOSS1244 protocluster is likely in an early evolutionary stage, far from virialization \citep{Shimakawa:2018ft,Shi:2021jt}.
During this phase, the funneling of pristine/low-metallicity cool gas through the filamentary structures can reach the local peaks of the underlying massive dark matter halos, where massive member galaxies reside \citep{Dekel:2006cn,2009MNRAS.395..160K}.
These cool gas streams thus dilute their metallicities and stimulate their star formation, consistent with our findings \citep[also see][]{Chartab:2021ed}.

\section{Conclusion and discussion}\label{sect:conclu}

We have presented hitherto the first measurement of the MZR in an overdense environment at $z\gtrsim2$ using grism slitless spectroscopy.
The grism data presented in this work are acquired by \mg, an \hst cycle-28 medium program to target three of the most massive galaxy protoclusters at $z\sim2.2-2.3$ identified using the MAMMOTH technique.
Using the complete \mg observations in the southwest region of the spectroscopically confirmed massive protocluster BOSS1244 at $z=2.24\pm0.02$,
we selected a sample of 36 protocluster member galaxies with $\Mstar/\Msun\in[10^9,10^{10.4}]$, $\mathrm{SFR}\in[10,240]$\,\Msun\,yr$^{-1}$, and $\oh\in[8.2, 8.7]$ assuming the B18 strong line calibrations.
We divide our galaxy sample into three mass bins: high ($\Mstar/\Msun\in[10^{10},10^{10.4})$), intermediate ($\Mstar/\Msun\in[10^{9.7},10^{10.0})$), and low ($\Mstar/\Msun\in[10^9,10^{9.7})$) containing 17, 11, and 8 galaxies respectively.
Our deep 3-orbit G141 grism exposures reach a 3-$\sigma$ SFR threshold of 9.2\,\Msun\,yr$^{-1}$, complete in sampling the $z\sim2.24$ SFMS down to the intermediate mass bin at $\Mstar\gtrsim10^{9.7}\Msun$.
The median SFR measurements in our intermediate(high) mass bin shows a $\sim$0.18(0.06) dex increase, as compared with the SFMS, revealing the environmental effect in boosting the SFR.
Applying our forward-modeling Bayesian metallicity inference method to the stacked spectra within the three mass bins, we derive the MZR in the BOSS1244 protocluster core as 
$\oh = \left(0.136\pm0.018\right) \times \log(\Mstar/\Msun) + \left(7.082 \pm 0.175 \right)$, assuming the B18 strong line calibrations.
After accounting for the SFR differences utilizing the FMR relation, we find that in comparison to the coeval galaxies of the same mass, our low-mass galaxies with $\Mstar/\Msun\in[10^9,10^{9.7})$ are more metal rich by $0.13\pm0.08$ dex (1.6$\sigma$), whereas our high-mass galaxies with $\Mstar/\Msun\in[10^{10},10^{10.4})$ are more metal poor by $-0.06\pm0.07$.
This is likely caused by the combined effect of efficient recycling of wind material ejected by galactic feedback in overdense environments and cold-mode gas accretion driven by massive cluster-scale dark matter halos. The former helps low-mass galaxies retain their metal production and leads to a metal enhancement, whereas the latter results in dilution of the metal content in high-mass galaxies, while both effects boost star formation in the protocluster environment.

With the ongoing data acquisition of our \mg program in the other two massive protocluster fields, \ie, BOSS1542 and BOSS1441 \citep{Cai:2017df,Shi:2021jt}, our statistics on protocluster member galaxies will greatly improve --- by a factor of 3 --- once all the data from the \mg program are acquired and analyzed. 
Furthermore, the true legacy of the \mg campaign will likely be the compilation of a unique sample of galaxies at the peak of cosmic star formation ($2\lesssim z\lesssim3$) in extremely overdense environments.
This unique galaxy sample, spectroscopically confirmed by \hst spatially-resolved grism spectroscopy, can
offer a valuable opportunity to explore the role that environment plays in a multitude of galaxy mass assembly processes, 
when followed up by further multi-wavelength observations.
For instance, with similar integration time as in the MOSDEF observations \citep{Kriek:2015ek}, Keck/MOSFIRE can provide ground-based K-band spectroscopy on our protocluster member galaxies with high spectral resolution to detect numerous emission lines (\eg, \Ha, \NII, \SII, \OI).
These data will be crucial in testing the environmental effects on the BPT offset, the electron density and excitation properties of the interstellar medium, and the dynamical masses inferred from the velocity dispersion of galaxies at $z\gtrsim2$ \citep{2015ApJ...801...88S,Sanders:2016ea,Price:2016gv}.
Keck/KCWI can probe the scenario of the cold-mode accretion through detection of \lya nebulae and interstellar absorption lines in the rest-frame ultraviolet \citep{Cai:2017ev,Cai:2019gz}.
While it is not feasible to obtain \Ha maps for our sample galaxies with \hst grisms, JWST is equipped with the NIRISS instrument which is capable of performing K-band (F200W) slitless spectroscopy.
Combined with our existing \mg G141 grism spectroscopy, the NIRISS/F200W exposures can offer spatially resolved dust maps and accurate star-formation maps at sub-kiloparsec scales, which are key in constraining the role of environment in the inside-out mass assembly of galaxies at cosmic noon \citep{Nelson:2015te,Nelson:2015vi}.
Multi-wavelength followup of these grism-selected protocluster member galaxies will open up new key windows on the exploration of environmental effects on galaxy evolution at high redshifts.

\acknowledgements
We would like to thank the anonymous referee for a careful read and useful comments that help improve the clarity of this paper.
This work is supported by NASA through HST grant HST-GO-16276.
We acknowledge the technical support from Tricia Royle and Norbert Pirzkal in scheduling our observations. XW is greatly indebted to Gabriel Brammer for his help in designing the observing strategy of this grism program, and his guidance in reducing the grism data. XW thanks Adam Carnall, Ranga-Ram Chary, Tucker Jones, and Ryan Sanders for useful discussion.
DDS and XZZ thank the support from the National Science Foundation of China (11773076 and 12073078), the National Key Research and Development Program of China (2017YFA0402703), and the science research grants from the China Manned Space Project with NO. CMS-CSST-2021-A02, CMS-CSST-2021-A04 and CMS-CSST-2021-A07. 

\facilities{HST (WFC3), LBT (LBC), CFHT (WIRCam)}.

\software{
\sex \citep{Bertin:1996hf},
\grzl \citep{Brammer:2021df},
\bagp \citep{Carnall:2018gb},
\adriz \citep{Hack:2021gv},
\tphot \citep{Merlin:2016uk},
LMFIT \citep{Newville:2021cv},
\emc \citep{ForemanMackey:2013io}.
}


\begin{longrotatetable}
{
\tabletypesize{\scriptsize}
\tabcolsep=2pt
\begin{deluxetable*}{cccccccccccccccc}   \tablecolumns{16}
\tablewidth{0pt}
\tablecaption{Measured properties of individual sources in the BOSS1244 protocluster member galaxy sample.}
\tablehead{
    \colhead{ID} & 
    \colhead{R.A.} & 
    \colhead{Dec.} & 
    \colhead{$z_\mathrm{grism}$} & 
    \multicolumn{2}{c}{\hst photometry [ABmag]} &
    \multicolumn{5}{c}{Observed emission line fluxes [$10^{-17}$\Funit]} &
    \multicolumn{5}{c}{Derived physical properties} \\
    & [deg.] & [deg.] &  & \multicolumn{2}{c}{\hrulefill}  &  \multicolumn{5}{c}{\hrulefill} & 
    \multicolumn{5}{c}{\hrulefill} \\
    &  &  &  & 
    \colhead{F125W} & 
    \colhead{F160W} & 
    \colhead{$f_{\rm [OII]}$} & 
    \colhead{$f_{\rm [NeIII]}$} & 
    \colhead{$f_{\rm H\gamma}$} & 
    \colhead{$f_{\rm H\beta}$} & 
    \colhead{$f_{\rm [OIII]}$} & 
    \colhead{$\log(M_{\ast}/M_{\odot})$} & 
    \colhead{A$_{\rm V}$} & 
    \colhead{SFR [\Msun/yr]} &
    \colhead{$12+\log({\rm O/H})^\mathrm{B18}$\tablenotemark{a}} & 
    \colhead{$12+\log({\rm O/H})^\mathrm{C17}$\tablenotemark{a}}
}
\startdata
    \multicolumn{16}{c}{high mass bin}   \\
    \noalign{\smallskip}
    00313 & 190.902894 & 35.851659 & 2.23 & 24.45 & 23.68 & $8.56\pm1.25$ & \nodata & $1.29\pm1.31$ & $4.50\pm0.78$ & $18.42\pm0.69$ & $10.31_{-0.08}^{+0.06}$ & $0.77_{-0.51}^{+0.69}$ & $52.15_{-24.27}^{+66.34}$ & $8.36_{-0.08}^{+0.08}$  &  $8.31_{-0.09}^{+0.07}$ \\
    00381 & 190.909141 & 35.854547 & 2.18 & 23.91 & 23.63 & $5.38\pm0.95$ & \nodata & $0.33\pm0.90$ & $1.56\pm0.63$ & $5.16\pm0.57$ & $10.03_{-0.20}^{+0.17}$ & $<0.40$ & $14.10_{-4.43}^{+12.04}$ & $8.51_{-0.09}^{+0.08}$ & $8.43_{-0.08}^{+0.07}$ \\
    00613 & 190.921119 & 35.865813 & 2.24 & 24.30 & 23.78 & $5.16\pm1.26$ & $0.38\pm2.23$ & $2.29\pm1.22$ & $2.05\pm0.82$ & $8.99\pm0.72$ & $10.18_{-0.10}^{+0.07}$ & $<0.79$ & $29.03_{-14.13}^{+50.89}$ & $8.40_{-0.12}^{+0.10}$ &  $8.33_{-0.11}^{+0.09}$ \\
    00769 & 190.909740 & 35.871348 & 2.24 & 23.43 & 23.05 & $15.72\pm2.34$ & $6.72\pm4.83$ & $1.17\pm2.80$ & $4.52\pm1.61$ & $24.76\pm1.84$ & $10.12_{-0.15}^{+0.11}$ & $<0.40$ & $48.91_{-14.52}^{+41.25}$ &  $8.38_{-0.08}^{+0.08}$ &  $8.33_{-0.07}^{+0.06}$ \\
    00996 & 190.873758 & 35.880726 & 2.32 & 23.96 & 23.41 & $2.82\pm0.89$ & $2.01\pm1.09$ & $1.31\pm0.79$ & $1.47\pm0.69$ & $7.81\pm0.73$ & $10.10_{-0.11}^{+0.09}$ & $<1.28$ & $<58.62$ & $8.29_{-0.15}^{+0.14}$ & $8.22_{-0.24}^{+0.12}$ \\
    01112 & 190.879330 & 35.886279 & 2.30 & 23.05 & 22.64 & $8.89\pm1.53$ & \nodata & \nodata & $4.46\pm1.16$ & $9.82\pm1.07$ & $10.13_{-0.05}^{+0.07}$ & $<0.55$ & $35.80_{-14.14}^{+42.45}$  &  $8.51_{-0.09}^{+0.08}$  &  $8.44_{-0.08}^{+0.07}$ \\
    01335 & 190.871757 & 35.895697 & 2.24 & 23.04 & 22.72 & $10.42\pm1.64$ & $3.06\pm3.35$ & $1.65\pm1.66$ & $2.70\pm0.79$ & $6.96\pm0.70$ & $10.13_{-0.09}^{+0.09}$ & $<0.25$ & $21.59_{-4.93}^{+10.12}$ & $8.56_{-0.08}^{+0.07}$ & $8.48_{-0.06}^{+0.05}$ \\
    01464 & 190.873403 & 35.901270 & 2.21 & 23.81 & 23.40 & $4.40\pm0.67$ & $2.05\pm0.88$ & \nodata & $2.32\pm0.44$ & $6.20\pm0.40$ & $10.08_{-0.12}^{+0.11}$ & $0.69_{-0.47}^{+0.64}$ & $21.83_{-9.52}^{+25.98}$ & $8.48_{-0.08}^{+0.07}$ & $8.41_{-0.08}^{+0.06}$ \\
    01998 & 190.868202 & 35.916375 & 2.21 & 22.83 & 22.49 & $10.84\pm2.36$ & \nodata & $2.20\pm2.70$ & $5.58\pm1.62$ & $22.81\pm1.45$ & $10.30_{-0.09}^{+0.08}$ & $0.94_{-0.62}^{+0.81}$ & $77.85_{-41.60}^{+127.41}$ & $8.36_{-0.12}^{+0.10}$ & $8.31_{-0.10}^{+0.08}$ \\
    02090 & 190.920739 & 35.918966 & 2.23 & 24.89 & 24.44 & $7.66\pm1.56$ & \nodata & $1.58\pm1.72$ & $4.14\pm0.85$ & $7.42\pm0.70$ & $10.03_{-0.10}^{+0.12}$ & $<0.53$ & $28.23_{-10.23}^{+30.81}$ & $8.61_{-0.08}^{+0.06}$ & $8.49_{-0.07}^{+0.06}$ \\
    02327 & 190.927963 & 35.926164 & 2.24 & 23.99 & 23.52 & $13.91\pm2.04$ & \nodata & $0.31\pm1.94$ & $5.18\pm1.41$ & $22.46\pm1.36$ & $10.01_{-0.15}^{+0.13}$ & $<0.48$ & $49.78_{-16.93}^{+45.18}$ & $8.38_{-0.08}^{+0.07}$ & $8.34_{-0.07}^{+0.07}$ \\
    02330 & 190.905581 & 35.926178 & 2.29 & 22.78 & 22.39 & $6.85\pm0.78$ & $0.56\pm1.01$ & $0.97\pm0.74$ & $5.81\pm0.60$ & $19.46\pm0.60$ & $10.34_{-0.07}^{+0.07}$ & $1.94_{-0.65}^{+0.54}$ & $238.55_{-129.30}^{+214.99}$ & $8.42_{-0.08}^{+0.07}$ & $8.36_{-0.11}^{+0.07}$ \\
    02588 & 190.872857 & 35.933017 & 2.24 & 23.18 & 22.78 & $9.85\pm1.39$ & $4.32\pm2.40$ & $1.84\pm1.44$ & $3.96\pm1.32$ & $9.34\pm1.33$ & $10.19_{-0.06}^{+0.08}$ & $<0.36$ & $32.68_{-9.83}^{+23.59}$ & $8.57_{-0.09}^{+0.07}$ & $8.47_{-0.08}^{+0.06}$ \\
    03061 & 190.838668 & 35.953087 & 2.23 & 23.07 & 22.73 & $12.00\pm1.37$ & \nodata & $0.69\pm1.27$ & $5.33\pm0.90$ & $19.14\pm0.80$ & $10.15_{-0.10}^{+0.10}$ & $<0.48$ & $45.24_{-15.46}^{+33.68}$ & $8.41_{-0.07}^{+0.07}$ & $8.36_{-0.07}^{+0.06}$ \\
    03155 & 190.842877 & 35.957764 & 2.32 & 23.48 & 23.02 & $6.76\pm1.31$ & \nodata & $1.38\pm1.41$ & $4.12\pm0.97$ & $9.49\pm0.90$ & $10.27_{-0.08}^{+0.09}$ & $<0.78$ & $49.74_{-24.87}^{+66.34}$ & $8.54_{-0.10}^{+0.08}$ & $8.44_{-0.09}^{+0.07}$ \\
    03276 & 190.857451 & 35.964185 & 2.21 & 22.48 & 22.18 & $17.43\pm1.79$ & $8.21\pm3.44$ & $1.58\pm1.88$ & $5.41\pm1.26$ & $24.53\pm1.09$ & $10.22_{-0.07}^{+0.10}$ & $<0.27$ & $44.17_{-9.88}^{+22.19}$ & $8.40_{-0.06}^{+0.06}$ & $8.35_{-0.06}^{+0.05}$ \\
    03495 & 190.853920 & 35.977367 & 2.23 & 23.72 & 23.22 & $7.72\pm1.13$ & \nodata & $1.99\pm1.04$ & $2.26\pm0.63$ & $10.71\pm0.57$ & $10.00_{-0.06}^{+0.09}$ & $<0.32$ & $20.79_{-5.14}^{+14.25}$ & $8.40_{-0.07}^{+0.07}$ & $8.35_{-0.07}^{+0.06}$ \\
    \noalign{\smallskip}\hline\noalign{\smallskip}
    \multicolumn{16}{c}{intermediate mass bin}   \\
    \noalign{\smallskip}
    01397 & 190.884983 & 35.898280 & 2.32 & 25.21 & 24.57 & $4.37\pm1.09$ & $0.49\pm1.49$ & \nodata & $0.71\pm0.74$ & $7.13\pm0.74$ & $9.74_{-0.11}^{+0.13}$ & $<0.77$ & $20.09_{-9.32}^{+38.07}$ & $8.33_{-0.12}^{+0.11}$ & $8.30_{-0.14}^{+0.09}$ \\
    01435 & 190.869813 & 35.900036 & 2.21 & 23.46 & 23.13 & $5.09\pm0.70$ & $1.32\pm0.84$ & $1.22\pm0.61$ & $2.72\pm0.46$ & $11.28\pm0.46$ & $9.89_{-0.07}^{+0.09}$ & $0.77_{-0.52}^{+0.61}$ & $30.90_{-14.27}^{+33.88}$ & $8.35_{-0.09}^{+0.08}$ & $8.31_{-0.09}^{+0.07}$ \\
    01467 & 190.869254 & 35.901455 & 2.21 & 24.00 & 23.51 & $2.94\pm0.51$ & $2.33\pm0.66$ & $1.02\pm0.45$ & $2.12\pm0.33$ & $10.62\pm0.33$ & $9.80_{-0.08}^{+0.08}$ & $1.22_{-0.77}^{+0.81}$ & $38.68_{-23.08}^{+61.41}$ & $8.24_{-0.10}^{+0.10}$ & $8.22_{-0.13}^{+0.10}$ \\
    01573 & 190.882538 & 35.905175 & 2.23 & 24.44 & 24.11 & $2.52\pm0.76$ & $1.68\pm1.03$ & \nodata & $1.86\pm0.78$ & $4.51\pm0.68$ & $9.92_{-0.24}^{+0.11}$ & $<0.92$ & $14.83_{-8.25}^{+46.41}$ & $8.35_{-0.17}^{+0.14}$ & $8.30_{-0.19}^{+0.11}$ \\
    01902 & 190.930795 & 35.913727 & 2.23 & 23.51 & 23.25 & $10.22\pm0.88$ & $1.16\pm1.32$ & $2.84\pm0.99$ & $4.54\pm0.72$ & $23.31\pm3.05$ & $9.83_{-0.06}^{+0.06}$ & $<0.44$ & $47.42_{-14.60}^{+34.02}$ & $8.31_{-0.08}^{+0.08}$ & $8.28_{-0.08}^{+0.07}$ \\
    01999 & 190.868601 & 35.916441 & 2.22 & 23.34 & 22.94 & $7.85\pm1.81$ & $7.95\pm3.35$ & $0.17\pm1.97$ & $4.35\pm1.29$ & $6.09\pm1.09$ & $9.75_{-0.09}^{+0.15}$ & $<0.57$ & $35.68_{-14.77}^{+46.62}$ & $8.66_{-0.10}^{+0.08}$ & $8.53_{-0.08}^{+0.07}$ \\
    02992 & 190.843993 & 35.949979 & 2.35 & 23.84 & 23.29 & $6.47\pm0.93$ & $0.65\pm1.35$ & $1.55\pm0.90$ & $3.71\pm0.74$ & $10.48\pm1.12$ & $9.95_{-0.06}^{+0.08}$ & $0.80_{-0.52}^{+0.60}$ & $49.02_{-23.26}^{+55.37}$ & $8.49_{-0.09}^{+0.08}$ & $8.40_{-0.09}^{+0.07}$ \\
    03385 & 190.853893 & 35.971115 & 2.30 & 24.05 & 23.53 & $4.75\pm0.87$ & \nodata & $1.29\pm0.88$ & $0.79\pm0.68$ & $5.14\pm0.65$ & $9.90_{-0.16}^{+0.16}$ & $<0.43$ & $14.70_{-4.77}^{+14.32}$ & $8.48_{-0.10}^{+0.09}$ & $8.40_{-0.08}^{+0.07}$ \\
    03480 & 190.870737 & 35.975923 & 2.15 & 23.02 & 22.77 & $10.53\pm1.58$ & $1.86\pm3.91$ & $3.98\pm1.62$ & $3.86\pm0.88$ & $11.55\pm0.86$ & $9.92_{-0.08}^{+0.09}$ & $<0.36$ & $29.19_{-8.12}^{+18.05}$ & $8.51_{-0.08}^{+0.07}$ & $8.42_{-0.07}^{+0.06}$ \\
 03496 & 190.846119 & 35.977685 & 2.24 & 23.79 & 23.47 & $7.51\pm1.34$ & \nodata & $0.78\pm1.28$ & $0.71\pm0.83$ & $4.95\pm0.81$ & $9.80_{-0.12}^{+0.09}$ & $<0.33$ & $13.94_{-3.94}^{+9.08}$ & $8.51_{-0.09}^{+0.08}$ & $8.46_{-0.07}^{+0.06}$ \\
 03516 & 190.851775 & 35.979114 & 2.23 & 23.79 & 23.46 & $5.49\pm1.23$ & $1.34\pm1.79$ & $1.01\pm1.15$ & $3.17\pm0.80$ & $9.89\pm0.79$ & $9.80_{-0.06}^{+0.10}$ & $0.97_{-0.64}^{+0.80}$ & $43.96_{-24.25}^{+68.05}$ & $8.45_{-0.11}^{+0.09}$ & $8.37_{-0.10}^{+0.08}$ \\
    \noalign{\smallskip}\hline\noalign{\smallskip}
    \multicolumn{16}{c}{low mass bin}   \\
    \noalign{\smallskip}
 00478 & 190.921015 & 35.858885 & 2.23 & 23.67 & 23.34 & $3.41\pm0.70$ & $1.29\pm0.78$ & $1.81\pm0.54$ & $1.38\pm0.36$ & $5.20\pm0.33$ & $9.48_{-0.16}^{+0.19}$ & $0.60_{-0.41}^{+0.62}$ & $13.06_{-5.10}^{+14.00}$ & $8.45_{-0.09}^{+0.08}$ & $8.37_{-0.09}^{+0.07}$ \\
 01070 & 190.886114 & 35.883869 & 2.30 & 24.43 & 24.07 & $7.21\pm1.32$ & $0.60\pm2.09$ & $0.69\pm1.24$ & $2.45\pm0.68$ & $17.81\pm0.60$ & $9.11_{-0.09}^{+0.10}$ & $<0.55$ & $22.71_{-8.09}^{+25.14}$ & $8.23_{-0.08}^{+0.08}$ & $8.23_{-0.09}^{+0.07}$ \\
 01071 & 190.867795 & 35.884012 & 2.24 & 25.11 & 24.36 & $3.62\pm0.96$ & \nodata & \nodata & $0.69\pm0.71$ & $5.41\pm0.63$ & $9.36_{-0.14}^{+0.17}$ & $<0.92$ & $18.21_{-9.97}^{+55.72}$ & $8.36_{-0.14}^{+0.12}$ & $8.32_{-0.15}^{+0.09}$ \\
 01394 & 190.877806 & 35.898194 & 2.21 & 23.81 & 23.61 & $5.19\pm0.61$ & $0.45\pm0.86$ & $1.07\pm0.53$ & $2.58\pm0.34$ & $6.86\pm0.31$ & $9.31_{-0.08}^{+0.11}$ & $0.50_{-0.32}^{+0.42}$ & $21.57_{-7.00}^{+14.06}$ & $8.51_{-0.06}^{+0.06}$ & $8.42_{-0.07}^{+0.06}$ \\
 01890 & 190.865939 & 35.913638 & 2.25 & 25.08 & 24.36 & $6.71\pm1.74$ & $0.66\pm2.52$ & $0.11\pm1.71$ & $1.77\pm1.22$ & $14.78\pm1.19$ & $9.14_{-0.06}^{+0.07}$ & $<0.92$ & $41.62_{-21.97}^{+92.18}$ & $8.28_{-0.13}^{+0.11}$ & $8.25_{-0.14}^{+0.09}$ \\
 02258 & 190.854129 & 35.924026 & 2.21 & 25.45 & 24.63 & $5.63\pm1.59$ & $2.65\pm2.26$ & \nodata & $1.87\pm0.81$ & $8.91\pm0.70$ & $9.54_{-0.12}^{+0.10}$ & $<1.14$ & $23.54_{-12.38}^{+55.02}$ & $8.35_{-0.14}^{+0.11}$ & $8.30_{-0.16}^{+0.09}$ \\
 02474 & 190.905843 & 35.929694 & 2.19 & 24.16 & 23.92 & $5.00\pm1.55$ & \nodata & $1.80\pm1.54$ & $1.40\pm1.01$ & $6.73\pm0.90$ & $9.39_{-0.06}^{+0.08}$ & $0.85_{-0.60}^{+1.14}$ & $24.38_{-12.85}^{+59.46}$ & $8.43_{-0.16}^{+0.13}$ & $8.35_{-0.18}^{+0.10}$ \\
 03331 & 190.862922 & 35.967686 & 2.22 & 24.04 & 23.71 & $4.12\pm0.99$ & \nodata & $0.13\pm0.90$ & $0.86\pm0.63$ & $7.89\pm0.59$ & $9.40_{-0.10}^{+0.12}$ & $<0.76$ & $18.67_{-8.69}^{+31.36}$ & $8.29_{-0.11}^{+0.10}$ & $8.27_{-0.13}^{+0.08}$ \\
    \noalign{\smallskip}\hline\noalign{\smallskip}
    \multicolumn{16}{c}{AGNs\tablenotemark{b}}   \\
    \noalign{\smallskip}
 00262 & 190.938355 & 35.849177 & 2.22 & 22.59 & 21.83 & $8.51\pm1.40$ & $0.88\pm2.69$ & \nodata & $1.69\pm1.01$ & $15.24\pm0.77$ & $11.43_{-0.04}^{+0.03}$ & $<0.46$ & $26.83_{-8.66}^{+27.59}$ & $8.31_{-0.08}^{+0.08}$ & $8.29_{-0.07}^{+0.07}$ \\
 00467 & 190.918225 & 35.858265 & 2.22 & 23.37 & 22.73 & $4.36\pm1.32$ & $5.97\pm2.67$ & $1.15\pm1.42$ & $3.26\pm0.94$ & $6.98\pm0.76$ & $10.98_{-0.07}^{+0.06}$ & $1.25_{-0.78}^{+1.01}$ & $53.21_{-32.90}^{+117.18}$ & $8.54_{-0.13}^{+0.09}$ & $8.44_{-0.13}^{+0.08}$ \\
 01165 & 190.897279 & 35.888298 & 2.35 & 24.17 & 23.34 & $5.99\pm1.07$ & $0.46\pm1.48$ & \nodata & $2.39\pm1.08$ & $21.10\pm1.69$ & $10.42_{-0.09}^{+0.11}$ & $<1.14$ & $58.87_{-31.93}^{+121.03}$ & $8.21_{-0.10}^{+0.10}$ & $8.18_{-0.10}^{+0.09}$
\enddata
    \tablecomments{The error bars shown in the table correspond to 1-$\sigma$ confidence intervals, whereas the upper/lower limits denote 2-$\sigma$ confidence limits. The table consists of four sections, corresponding to the three mass bins selected for stacking, and the AGN sample. The high, intermediate, and low mass bins are defined as $\log(\Mstar/\Msun) \in$~[10.0, 10.4), [9.7, 10.0), and [9.0, 9.7), respectively (see Sect.~\ref{subsect:stack}).}
    \tablenotetext{a}{The superscript indicates the sets of strong line calibrations (B18: \citet{Bian:2018km}, C17: \citet{Curti:2017fn}) used in the metallicity inference, with the coefficients presented in Table~\ref{tab:coef}.
    We consider the results based on the B18 calibrations as our default results, for the sake of a direct comparison with the field measurements in \citet{Sanders:2021ga} derived using the same set of calibrations.}
    \tablenotetext{b}{For sources in this section, their \oh estimates are not trustworthy since their nebular emissions are dominated by AGN ionization and therefore strong line calibrations are no longer applicable.}
\label{tab:indvd}
\end{deluxetable*}
}

\end{longrotatetable}

\bibliographystyle{apj}
\bibliography{bibtexlib}

\end{document}